\documentclass[
reprint,
superscriptaddress,
amsmath,amssymb,
aps,
pra,
floatfix
]{revtex4-2}

\usepackage{graphicx}
\graphicspath{ {./} }
\usepackage{dcolumn}
\usepackage{bm}
\usepackage{hyperref}

\usepackage{braket}
\usepackage{relsize}
\usepackage{extarrows}
\usepackage[skins]{tcolorbox}
\usepackage[framemethod=tikz]{mdframed}

\begin{document}

\preprint{APS/123-QED}

\title{Using Quantum Switches to Mitigate Noise in Grover's Search Algorithm}

\author{Suryansh Srivastava}
 \altaffiliation[Corresponding Author]{}
\email{suryansh.srivastava@research.iiit.ac.in}
\affiliation{%
 Centre for Quantum Science and Technology,\\ International Institute of Information Technology Hyderabad, Gachibowli, Hyderabad-500032, Telangana, India.
}
\affiliation{
 Center for Computational Natural Sciences and Bioinformatics,\\ International Institute of Information Technology, Gachibowli, Hyderabad-500032, Telangana, India.
}

\author{Arun K. Pati}
\email{akpati@iiit.ac.in}
\affiliation{
Centre for Quantum Engineering, Research and Education (CQuERE), TCG CREST, Salt Lake, Sector 5, Kolkata-700091, India
}
\affiliation{%
 Centre for Quantum Science and Technology,\\ International Institute of Information Technology Hyderabad, Gachibowli, Hyderabad-500032, Telangana, India.
}
\author{Indranil Chakrabarty}
\email{indranil.chakrabarty@iiit.ac.in}
\affiliation{%
 Centre for Quantum Science and Technology,\\ International Institute of Information Technology Hyderabad, Gachibowli, Hyderabad-500032, Telangana, India.
}
\affiliation{
 Center for Security, Theory and Algorithmic Research,\\ International Institute of Information Technology Hyderabad, Gachibowli, Hyderabad-500032, Telangana, India.
}%

\author{Samyadeb Bhattacharya}
\email{samyadeb.b@iiit.ac.in}
\affiliation{%
 Centre for Quantum Science and Technology,\\ International Institute of Information Technology Hyderabad, Gachibowli, Hyderabad-500032, Telangana, India.
}
\affiliation{
 Center for Security, Theory and Algorithmic Research,\\ International Institute of Information Technology Hyderabad, Gachibowli, Hyderabad-500032, Telangana, India.
}%



\date{\today}

\begin{abstract}
Grover's quantum search algorithm promises a quadratic speedup for unstructured search over its classical counterpart. But this advantage is affected by noise acting on the search space. Here, we show that a quantum switch can act as a resource to mitigate the effects of noise. In this scenario, the noise is modeled by a depolarizing channel, which coherently acts on the entire quantum register. 
We show that a quantum switch can significantly reduce the error in Grover's search algorithm. We consider the success probability of finding the marked item as the sole quantifier of diminishing the effect of noise in the search space in the presence of quantum switch. We propose two frameworks for the application of quantum switches. In the first framework, we apply the superposition of channel's orders in the form of a switch and do a post-selection at every iteration of the applications of the Grover operator.
In the second framework, we delay this measurement and post-selection until the very end. 
The number of post selections is minimal in the second scenario, and hence the noise reduction can be attributed more to the presence of quantum switch. 
We illustrate with an example of significant advantage in the success probability of Grover's algorithm using quantum switch.\\

\end{abstract}

\maketitle
\section{Introduction}
In the current landscape of quantum computing, where innovative algorithms are being introduced more regularly, there's a shift in focus from the anticipation of a fault-tolerant quantum computer towards the reality of noisy intermediate scale quantum (NISQ) devices \cite{preskill_quantum_2018,bharti_noisy_2022}. Nevertheless, fundamental and pioneering algorithms like Shor's algorithm \cite{shor_polynomial-time_1997}, Deutsch's algorithm \cite{deutsch_d_rapid_1992}, and Grover's search algorithm \cite{grover_fast_1996} introduced in the late part of the previous century are becoming more and more relevant. Grover’s search algorithm dates back to the early discoveries of quantum algorithms, which leverages the phenomenon of quantum superposition to obtain a quadratic speedup in searching a desired element in an unstructured database \cite{grover_fast_1996}. This gives us a significant advantage over the classical unstructured search, which is essentially linear and can best be optimized to $\mathcal{O}(N)$ steps for a database of $N$ elements. The Grover search algorithm was designed to search through the same space in $\mathcal{O}(\sqrt{N})$ steps.
The main idea of the algorithm is to amplify the probability amplitudes of the states that represent the target elements based on a selection function (oracle) while also reducing the probability amplitudes of other states by inversion around the mean.
We reach the target states in $(\pi/4)\sqrt{N}$ steps with a high probability of success. 
Many variations of the Grover search algorithm were introduced \cite{ambainis_quantum_2004}. This includes features like a dynamic selection function in contrast to the static selection function by enabling a recommendation algorithm \cite{chakrabarty_dynamic_2017}. 
Grover's search algorithm has wide-ranging applications in various domains, such as statistics, like extracting the minimum element \cite{durr_quantum_1999}, computing quantities representing averages like mean\cite{tucci_quantum_2014} and median \cite{grover_median_1996} from an unordered data set more quickly than is possible on a classical computer, and computational It enhances the speed of solving NP-complete problems \cite{furer_solving_2008}, such as those involving exhaustive searches \cite{ambainis_quantum_2005}. It also improves the efficiency of general constraint satisfaction problems \cite{cerf_nested_2000}. In quantum computing, Grover's algorithm is used for solving black-box problems like element distinctness \cite{ambainis_quantum_2005} and collision problems \cite{brassard_quantum_1998}, where it helps in quickly finding specific inputs from known outputs in functions. This makes it a valuable tool in cryptography, e.g., for breaking encryption through brute-force attacks, like collision attack \cite{bernstein_post-quantum_2009}, and for a quantum attack on block ciphers, leading to the use of the Grover oracle for key search in AES \cite{jaques_implementing_2019}. The concept of Grover coin/Grover walk was introduced and used in random walks to show advantages \cite{mandal_limit_2022}.\\ 

\noindent Studying the role of the environment in the evolution of quantum systems is essential as it helps us to understand the effect of noise on these resourceful states. These environmental interactions can be modeled as quantum channels, cf.(\ref{subsec:noise_modeling_depolarising_channel})

Investigations were made in the context of the effect of noise on various quantum properties, starting from entanglement \cite{braunstein_speed-up_2002, chakraborty_entanglement_2013} to the broader aspect of correlation \cite{chakraborty_non-classical_2013}. These effects were studied not only from the perspective of a closed but also from an open system viewpoint\cite{BRE02}. In particular, 
studies were reported on the effects of the depolarizing channel as a manifestation of noise in Grover's search algorithm \cite{cohn_grovers_2016, vrana_fault-ignorant_2014}. This algorithm's efficacy is notably compromised when confronted with scenarios involving a non-ideal oracle \cite{regev_impossibility_2008,temme_runtime_2014} or in the presence of environmental quantum noise \cite{chen_searching_2003}. This degradation in performance highlights the sensitivity of the algorithm's operational efficiency to external perturbations and the integrity of the oracle mechanism.\\

\noindent In contemporary quantum information research, there has been a growing interest in exploring the concept of indefinite causal order in quantum systems and its potential as a resource in various information processing tasks. Originally proposed by L. Hardy \cite{ICO_1, ICO_2} and practically applied in information theory by Chribella et al. \cite{chiribella_quantum_2013}, the concept led to the development of a quantum switch that uses an ancillary system to control the sequence of two quantum operations, $E1$ and $E2$, on a quantum state $\rho$, making the order indefinite. Oreshkov et al. \cite{oreshkov_quantum_2012} furthered this concept, employing process matrix formalism to establish a more robust framework of causal indefiniteness. This innovation has found applications in various fields, including quantum channel probing \cite{chiribella_perfect_2012}, nonlocal games \cite{oreshkov_quantum_2012}, quantum metrology \cite{zhao_quantum_2020}, quantum communication \cite{ebler_enhanced_2018, salek_quantum_2018, chiribella_indefinite_2021, koudia_how_2022}, quantum internet \cite{caleffi_quantum_2020} as well as in reducing quantum communication complexity \cite{guerin_exponential_2016}, and advancing quantum computing \cite{araujo_computational_2014, simonov_universal_2023} and thermodynamics \cite{guha_thermodynamic_2020}. Its most recent application has transformed absolutely separable states into resourceful states. \cite{yanamandra_breaking_2023}. The practicality and advantages of indefinite causal order have also been confirmed through recent experiments \cite{procopio_experimental_2015, rubino_experimental_2017, goswami_indefinite_2018}.\\

\noindent Despite the difficulties presented by noise in the quantum search algorithm, some approaches have been proposed to retain its quantum advantage \cite{saha_asymptotically_2022, avron_quantum_2021, maciejewski_mitigation_2020}. In this work, we explore another approach to preserve this advantage by using a quantum switch to potentially reduce the detrimental effects of noise, modeled by a depolarizing channel, on the success probability of the Grover search algorithm.
We will use the success probability as a metric to understand how a switch significantly reduces the effect of noise. In other words, it enables us to tolerate more error in the circuit before losing its advantage over its classical analog. Here, we consider two particular frameworks where, in one framework (\ref{subsec: F1_xi}), we measure at the end of each iteration to trace out the switch and compare Grover's success probability in a noisy scenario to show a slight advantage. In the other framework \ref{subsec: F2_omega}, we consider a register of switches and allow the input state to undergo each iteration with different switches applied till the end instead of measuring in between. The second approach preserves the advantage significantly better than
the first model in a noisy scenario. Our result shows that the switch can significantly mitigate the effect of the noise that builds up in the algorithm and preserve its efficacy for longer.\\
\noindent The rest of the paper is organized as follows. In section \ref{sec:related_concepts}, we describe relevant concepts like the Grover search algorithm, modeling the depolarizing channel as noise, and quantum switch. In section \ref{sec:noisy_grover_with_depolarising_noise}, we describe the role of noise in reducing the success probability of Grover's search algorithm. In section \ref{sec:application_q_switch_grovers}, we consider the effect of the switch in enhancing the success probability on the platform of two different frameworks. $^{3.3}$Lastly, section \ref{sec:conclusions} contains conclusions and summary of results  and later appendices \ref{sec:appendix} contain detailed calculations from various portions of the paper.

\section{Related Concepts}\label{sec:related_concepts}
This section discusses a few relevant concepts that are prerequisites before discussing our manuscript's main findings. 

\subsection{Grover's Search Algorithm}

\begin{figure}[b]
\noindent \begin{mdframed}
[
        linecolor=black,linewidth=0.5pt, frametitlerule=true,
        apptotikzsetting={\tikzset{mdfframetitlebackground/.append style={
            shade,left color=white, right color=gray!20}}}, 
        frametitlerulecolor=black,
        frametitlerulewidth=0.5pt, innertopmargin=\topskip,
        frametitle={Grover Search Algorithm},
]
\includegraphics[scale=0.225]{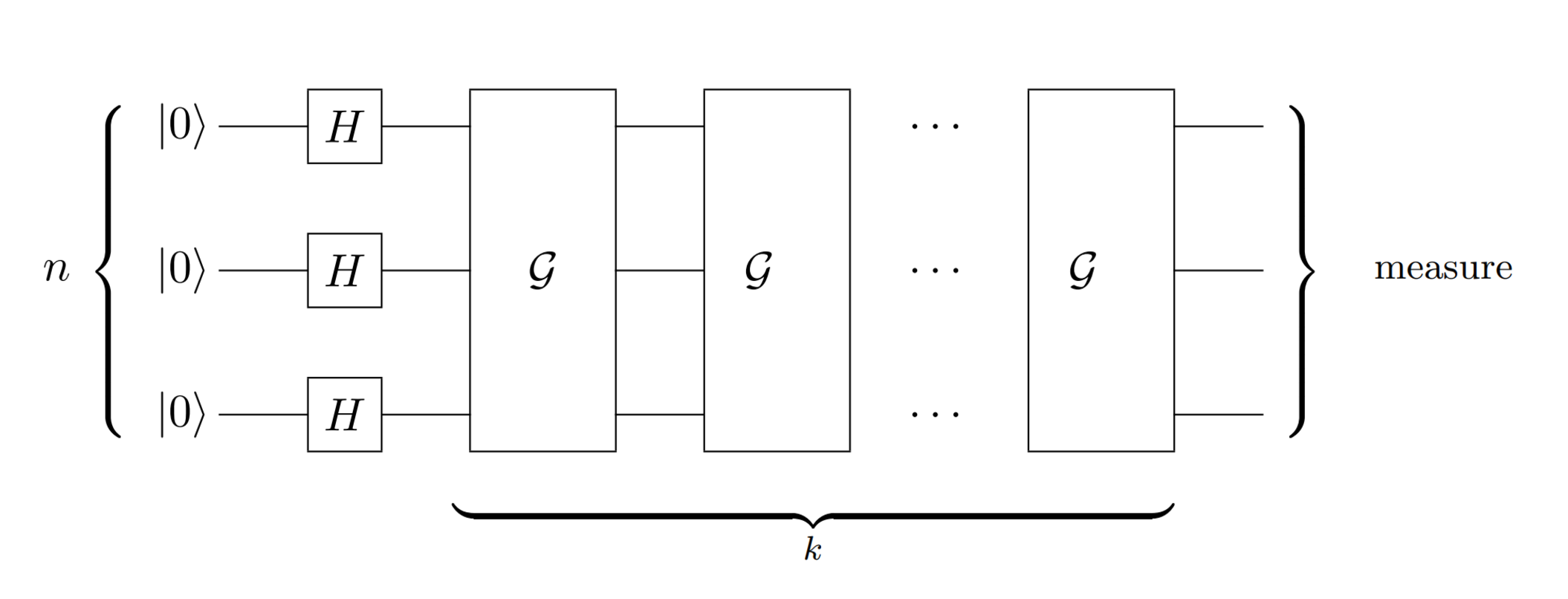}
\caption{\label{fig:ideal_grovers} Grover’s algorithm, with k Grover iterates}
\end{mdframed}
\end{figure}

Grover's search algorithm, a quantum algorithm discovered in the early stages of quantum \linebreak computing, revolutionized the search process of an unstructured database. It was proved to be optimal 
to solve the problem of finding a target element in an unsorted database of $N$ elements using $k_{Gr}=\left \lfloor\frac{\pi}{4} \sqrt{N}\right\rfloor$ oracle queries \cite{zalka_grovers_1999, bennett_strengths_1997}. The unstructured search problem can be described as follows: Consider a search space containing $N = 2^{n}$ quantum states, where $n$ represents the number of qubits defining the Hilbert space $\left(\mathcal{H} = \mathbb{C}^{2^{n}}\right)$.
Within this database, an unknown target state exists. Provided with an oracle (or a black box) that can confirm whether a selected element is the target, the objective is to identify this target state with a high probability of success while minimizing the number of steps required in the process.\\

\noindent \textbf{Algorithm:} 
To simplify things, let us assign indices $[0, N-1]$ to each element in the unstructured database, consisting of $N$ elements. Here, we assume $N = 2^n$, with $n \in \mathbb{Z}_{+}$ being the number of qubits in the register, and each possible quantum state of the register is used to encode each entry in the database. We denote these states as $\{|0\rangle,|1\rangle,\ldots|N-1\rangle\}$. 
We introduce a selection function $f$, which takes a state $|x\rangle$ as an input, $x \in [0, N - 1]$, and assigns a value $1$ when the state meets the search criteria and $0$ otherwise. This function can be understood as an unstructured database query, where the domain represents indices in the database.
\begin{equation}
f = \left\{ \begin{array}{cc}
    0  & \textit{if x is not chosen}, \\ 
    1 & \textit{if x is chosen.}
\end{array}\right.
\end{equation}
\noindent Let $\ket{\tau}$ denote the target state. The primary objective of the algorithm is to find this particular index \(\tau\).
To interact with the function \( f \), Grover's algorithm uses a subroutine, often referred to as an oracle or a black box function, 
represented by a unitary operator \( O_{\tau} \). 
The action of the operator on the computational basis is given by,

\[ O_{\tau}|x\rangle = (-1)^{f(x)}|x\rangle. \]

\noindent The algorithm is shown in fig (\ref{fig:ideal_grovers}). 
In the first step, we create a superposition of $N$ quantum states. The application of a Hadamard transformation ($H$) does this,
\begin{equation}
    \ket{s} = \frac{1}{\sqrt{N}} \sum_{x=0}^{N-1}|x\rangle.
\end{equation}
After that, the algorithm dictates the repeated application of a quantum subroutine referred to as the Grover iteration or as the Grover operator $\mathcal{G}$.\\

\noindent Each \textbf{Grover iteration} $\mathcal{G}$ consists of following steps:
\begin{itemize}

    \item Apply the oracle operator $O_{\tau}=2|\tau\rangle\langle \tau|-\mathbb{I}$.
    \item Apply the diffusion operator \(\delta\) that performs a rotation around the state \(
    \ket{s}\). The operator \(\delta\) itself can be formulated as:
\[ \delta = -H^{\otimes n}(2|0\rangle\langle 0| - I)H^{\otimes n} = 2|s\rangle\langle s| - \mathbb{I}\]

    \item Perform measurements in the canonical basis in each qubit. The target state will emerge with a high probability 
\end{itemize}

After applying the oracle operator and diffusion operator $k$ times, the result is
\begin{equation}
\rho(k)= \mathcal{G}^{k} \rho(0)\left(\mathcal{G}^{\dagger}\right)^{k},\label{eqn: ideal_grover_state_k}
\end{equation}
\noindent where $\mathcal{G}=\delta O_{\tau}$ and $\rho(0) = \ket{s}\bra{s}$. It can be shown that the density operator obtained afterward is $\rho(k)=\left|s_{k}\right\rangle\left\langle s_{k}\right|$, where
\begin{eqnarray}
\left|s_{k}\right\rangle &&=\sin\left((2k+1)^{3.4.b}\arcsin\left(\frac{1}{\sqrt{N}}\right)\right)|\tau\rangle\nonumber\\
&&+\cos\left((2k+1)\arcsin\left(\frac{1}{\sqrt{N}}\right)\right)|1-\tau\rangle, {}\nonumber\\
&&\text{where, }|1-\tau\rangle =\frac{1}{\sqrt{N-1}} \sum_{\substack{i=0 \\
i \neq \tau}}^{N-1}|i\rangle
\end{eqnarray}
\noindent Hence, the probability of success of finding the target element after $k$ steps is
\begin{equation} \label{ideal_grover_success_probability}
P(k)=\sin ^{2}\left((2 k+1) \arcsin\left(\frac{1}{\sqrt{N}}\right)\right). 
\end{equation}

\subsection{Noise modeled as Quantum Depolarizing Channel}\label{subsec:noise_modeling_depolarising_channel}
A quantum channel characteristically represents a completely positive trace-preserving (CPTP) map and a convex linear transformation, facilitating the transition of quantum states within a quantum system \cite{nielsen_quantum_2010}.
A quantum depolarizing channel serves as a model for quantum noise in quantum systems. In this model, the $d$-dimensional depolarizing channel is conceptualized as a CPTP map, \(\Lambda_{q}\), which is characterized by a single parameter, $q$ \cite{wilde_quantum_2017}. This map transforms a $d$-dimensional quantum state $\rho$ into a mixture of the original state and the maximally mixed state in $d$-dimensions, $\frac{\mathbb{I}_d}{d}$, mathematically represented as:
\begin{equation}\label{def:general_depolarising}
\Lambda_{q}(\rho) = q\rho + \frac{(1-q)}{d}\mathbb{I}_d.
\end{equation}

For \(\Lambda_{q}\) to maintain complete positivity, the parameter \(q\) must adhere to the following constraint \cite{das_quantum_2019}:
\begin{equation}
\frac{-1}{d^2-1} \leq q \leq 1. 
\end{equation}
A subset of the generalized qudit depolarizing channel as defined in (\ref{def:general_depolarising}), denoted as $\mathcal{D}_t$, is a specific type of quantum channel applicable to a $d$-dimensional quantum state $\rho$. For a parameter $0 < t < 1$, it modifies $\rho$ into the channel output 

\begin{equation}
\mathcal{D}_t(\rho) = (1 - t)\frac{\mathbb{I}_d}{d} + t\rho.
\end{equation}

The parameter $1 - t$ represents the channel depolarising probability, whereas $t$ signifies the likelihood of retaining the state $\rho$. This model of the depolarizing channel is frequently employed to represent quantum noise \cite{ebler_enhanced_2018}, given by,

\begin{eqnarray}
\mathcal{D}_t(\rho) = t\rho+\frac{1-t}{d^{2}} \sum_{i=1}^{d^{2}} U_{i} \rho U_{i}^{\dagger}.
\end{eqnarray}
Here $\left\{U_{i}\right\}_{i=1}^{d^{2}}$ are unitary operators and form an orthonormal basis of the space of $d \times d$ matrices. This is the notation we will use throughout the rest of the paper.
Quantum channels have operator-sum representations in terms of Kraus operators. As per the operator-sum representation \cite{nielsen_quantum_2010}, the depolarizing channel in a $d$-dimensional space can be written as:
\begin{equation} \label{eqn:kraus_operator_D_t}
    \mathcal{D}_t (\rho) = \sum_{i=0}^{d^2} D_i \rho  D_i^\dagger,
\end{equation}
where the Kraus operators $D_i$ are defined as $D_0 = \sqrt{t} \mathbb{I}_d$ and $D_i = \frac{(1 - t)}{\sqrt{d}} U_i$ for $i = 1, \ldots, d^2$ cf.(\ref{eqn: kraus_unitary_operators_calc}). Here, $U_i$ represents orthogonal unitary operators. 


\noindent One notable property of this channel is its commutative nature under sequential applications. Utilizing two such channels, $\mathcal{D}_{t_1}$ and $\mathcal{D}_{t_2}$ with preservation probabilities $t_1$ and $t_2$, respectively, regardless of the order of application, results in another depolarizing channel with a combined state preservation probability of $t_1t_2$:

\begin{equation}
\mathcal{D}_{t_2}(\mathcal{D}_{t_1}(\rho)) = \mathcal{D}_{t_1}(\mathcal{D}_{t_2}(\rho)) = (1 - t_1t_2)\frac{\mathbb{I}_d}{d} + t_1t_2\rho.
\end{equation}
\noindent In many cases, this channel model might be overly pessimistic. Often, insights into the physical characteristics of the channel can be gleaned through various estimation methods. The deployment of the depolarizing channel as a noise model should be considered mainly when there is a lack of detailed information about the nature of the actual physical channel and can be represented as follows:
\begin{equation}\label{eqn:noisy_grover_with_TPDCh}
    \mathcal{D}_t(\rho) := (1-t)\rho_d + t\frac{\mathbb{I}_d}{d}\mathrm{Tr}[\rho_d],
\end{equation}
\noindent acting on states $\rho$ on a d-dimensional Hilbert space.
In some situations, it is convenient to think about quantum channels as defined in the space of all linear operators on a given Hilbert space, not just density matrices. This will be the case for some calculations in \ref{appendix: switched_kgt1}. The $\mathrm{Tr}[\rho]$ factor is necessary to ensure that $\mathcal{D}_t$ is linear. If the input is known to have a unit trace, then the $\mathrm{Tr}[\rho]$ factor is not necessary.\\

The depolarizing channel's significance is evident from its historical role in the study of channel identification \cite{martinez_superadditivity_2023}, its application in enhancing channel capacity through indefinite causal order \cite{chiribella_indefinite_2021}, and its use in analytical comparisons of various channel probing methodologies \cite{frey_probing_2011}. 

\subsection{Quantum Switch}
\begin{figure}[ht]
\noindent \begin{mdframed}
[
        linecolor=black,linewidth=0.5pt, frametitlerule=true,
        apptotikzsetting={\tikzset{mdfframetitlebackground/.append style={
            shade,left color=white, right color=gray!20}}}, 
        frametitlerulecolor=black,
        frametitlerulewidth=0.5pt, innertopmargin=\topskip,
        frametitle={Quantum Switch},
]

\includegraphics[scale=0.49]{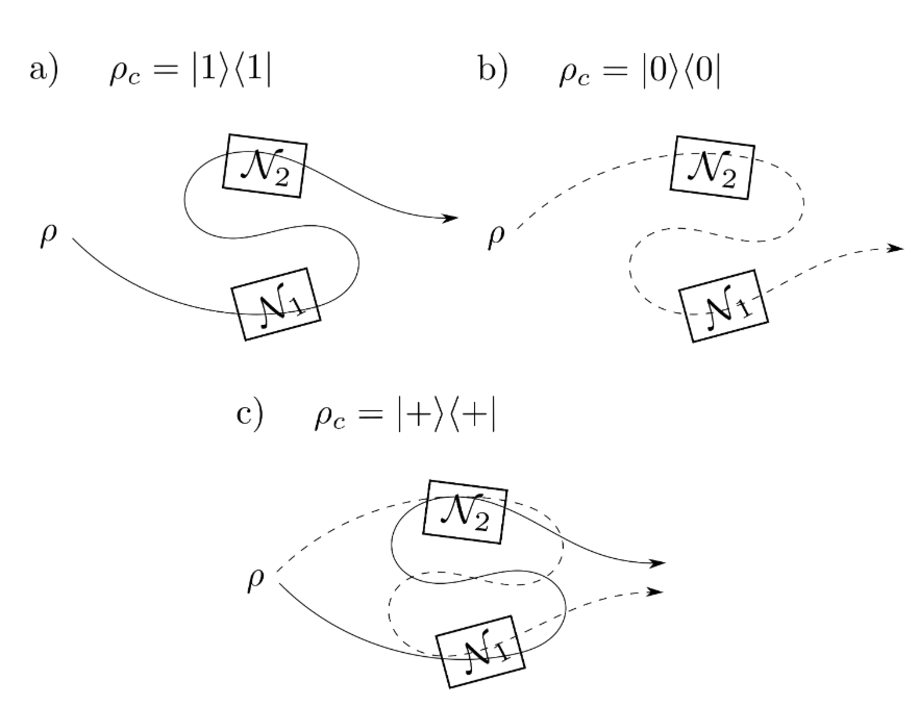}
\caption{\label{fig:indefinite_causal_channel-order} Fixed order vs superposition of orders. (a) A quantum particle, prepared in the state $\rho$, goes first through channel $\mathcal{N}_1$ and then through channel $\mathcal{N}_2$. This configuration is associated with the state $\rho_c = \ket{1}\bra{1}$ of a control qubit, in which the choice of order is encoded. (b) The quantum particle goes first through $\mathcal{N}_2$ and then through $\mathcal{N}_1$. This alternative configuration is associated with the qubit state $\rho_c = \ket{0}\bra{0}$. (c) The quantum switch creates a superposition of the two configurations (a) and (b). It takes a control qubit in a superposition state, such as $\rho_c = \ket{+}\bra{+}$, and correlates the order of the two channels with the state of the qubit.}
\end{mdframed}
\end{figure}

In this subsection, we discuss the implementation of quantum switches in general. As mentioned before in eq.(\ref{eqn:kraus_operator_D_t}), the action of a quantum channel \(\mathcal{N}\) on an input state \(\rho\) can be expressed using the Kraus representation, or operator-sum representation, formulated as \(\mathcal{N}(\rho) = \sum_i K_i \rho K_i^\dagger\), where \(\{K_i\}\) denotes the set of Kraus operators for \(\mathcal{N}\). Consider two quantum channels, \(\mathcal{N}_1\) and \(\mathcal{N}_2\); they can operate either concurrently or sequentially. The concurrent operation is represented as \(\mathcal{N}_1 \otimes \mathcal{N}_2\). In contrast, sequential operations can be arranged in two ways: \(\mathcal{N}_1\) followed by \(\mathcal{N}_2\) (notated as \(\mathcal{N}_2 \circ \mathcal{N}_1\)) or \(\mathcal{N}_2\) followed by \(\mathcal{N}_1\) (notated as \(\mathcal{N}_1 \circ \mathcal{N}_2\)). If the sequence of these channels is fixed, only one of the sequences, either \(\mathcal{N}_2 \circ \mathcal{N}_1\) or \(\mathcal{N}_1 \circ \mathcal{N}_2\), is permissible. However, 
the sequence in which two channels operate can be rendered indefinite using an ancillary system, namely the control qubit (\(\rho_c\)) \cite{ebler_enhanced_2018, bhattacharya_random-receiver_2021, chiribella_indefinite_2021}, \(\rho_c = \ket{c}\bra{c}\) where \(\ket{c} = \sqrt{\theta}\ket{0}+\sqrt{\overline{\theta}}\ket{1}\). With \(\rho_c\) set to the \(|0\rangle \langle 0|\) state, the \(\mathcal{N}_2 \circ \mathcal{N}_1\) arrangement affects the state \(\rho\), while \(\mathcal{N}_1 \circ \mathcal{N}_2\) comes into play when \(\rho_c\) is in the \(|1\rangle \langle 1|\) state. If \(\{K^{(1)}_i\}\) represent the Kraus operators for \(\mathcal{N}_1\) and \(\{K^{(2)}_j\}\) for \(\mathcal{N}_2\), the generalized Kraus operator can be expressed as 
\begin{equation}\label{def:switch_kraus_0}
    \mathcal{W}_{ij} = K^{(2)}_j \circ K^{(1)}_i \otimes |0\rangle\langle 0| + K^{(1)}_i \circ K^{(2)}_j \otimes |1\rangle\langle 1|,
\end{equation}
where ${K}^{(1)}_i$ is the $i^{th}$ Kraus operator of channel $\mathcal{N}_1$ and ${K}^{(2)}_j$ is the $i^{th}$ Kraus operator of channel $\mathcal{N}_2$; and $|0\rangle\langle0|_{c}, 1\rangle\langle1|_{c}$ are the projectors associated with the native basis of $|c\rangle$. The cumulative evolution of the combined system is thus described by
\begin{equation}
S(\mathcal{N}_1, \mathcal{N}_2) (\rho \otimes \rho_c) = \sum_{i, j} \mathcal{W}_{ij}(\rho \otimes \rho_c)\mathcal{W}_{ij}^\dagger. 
\end{equation}
This arrangement, in which the order of application of the two channels is coherently controlled, captures the operation of the quantum switch. It can be thought of as a higher order map that takes two channels $\mathcal{N}_{1}, \mathcal{N}_{2}$ as inputs and gives the superposition of their orders as outputs depending upon the choice of the control qubit $\ket{c}$.
Upon completion, the control qubit is measured in the basis \(\{|+\rangle, |-\rangle\}\) where \(\ket{\pm} = \frac{\ket{0} \pm \ket{1}}{2}\). For each measurement outcome, the corresponding reduced state of the target qubit is obtained as 
\begin{equation}
    \sideset{_c}{}{\mathop{\bra{\pm}}}\mathcal{S}(\mathcal{N}_1, \mathcal{N}_2)(\rho \otimes \rho_c)\ket{\pm}_c.
\end{equation}
This encapsulates the functioning of the Quantum switch, as depicted schematically in fig.(\ref{fig:indefinite_causal_channel-order}).

\section{Noise in Grover's Search Algorithm}\label{sec:noisy_grover_with_depolarising_noise}
Quantum systems inevitably interact with their environment, introducing errors and noise when these systems are externally controlled, such as during gate applications or state preparation. This influence on quantum algorithms has been the subject of a recent investigation by multiple researchers \cite{ambainis_grovers_2013, azuma_decoherence_2002}. Quantum search algorithm constitutes an oracle-based method designed to search an unordered database efficiently. This algorithm showcases a quadratic speedup compared to the classical brute-force search \cite{grover_fast_1996}. Nonetheless, this advantage significantly diminishes when the oracle encounters faults \cite{temme_runtime_2014} or when the algorithm operates in an environment of noise \cite{salas_noise_2008}. Prior research was conducted by Vrana et al. on partial depolarising and dephasing noise models \cite{vrana_fault-ignorant_2014}. Apart from the total depolarising error model, which acts on the entire quantum register simultaneously, Cohn et al. \cite{cohn_grovers_2016} also studied the local depolarizing error model that acts on each qubit of the register separately with a certain probability and holds significance as it reflects the impact of gate errors on any practical implementation of a quantum circuit.

\begin{figure*}[htp]
\noindent \begin{mdframed}
[
        linecolor=black,linewidth=0.5pt, frametitlerule=true,
        apptotikzsetting={\tikzset{mdfframetitlebackground/.append style={
            shade,left color=white, right color=gray!20}}}, 
        frametitlerulecolor=black,
        frametitlerulewidth=0.5pt, innertopmargin=\topskip,
        frametitle={Noisy Grover's Search Algorithm},
]

\includegraphics[scale=0.6]{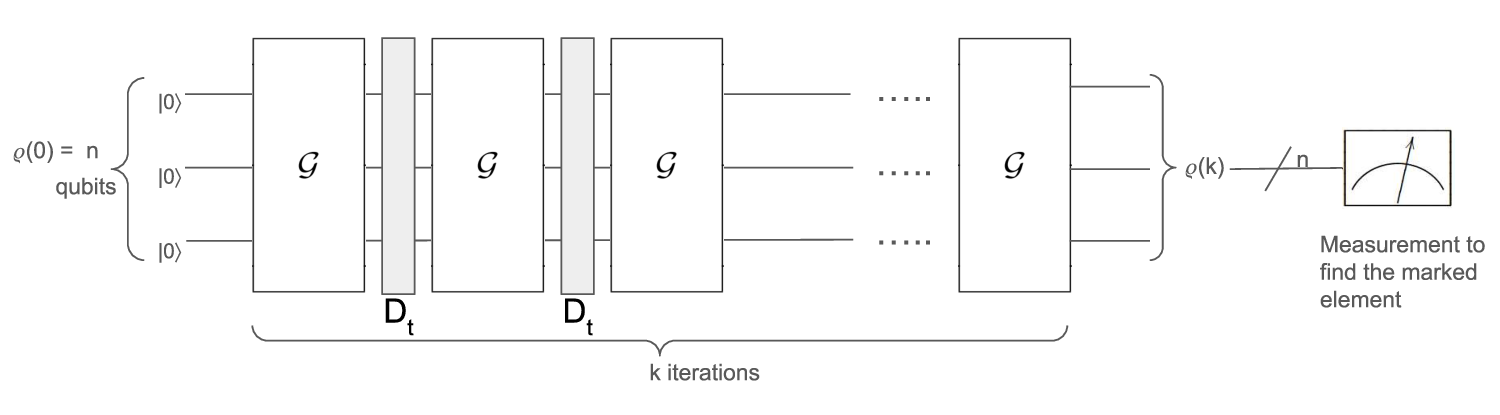}
\caption{\label{fig:noisy_grover_with_TPDCh} In this figure, we depict the action of noise on the Grover operator $\mathcal{G}$. They are represented by the shaded bar. Here, we are measuring at the end to see the effect of noise on the Grover operator's success probability. }
\end{mdframed}
\end{figure*}
\noindent In this section, we briefly discuss the impact of a known noise rate on Grover’s search algorithm.
The resultant errors manifest between the execution of two quantum gates can be effectively modeled using depolarizing channels \(\mathcal{D}_t\). This modeling approach simulates continuous noise exposure in quantum computers and the stochastic nature of error introduction. Depolarizing noise discards the whole quantum register consisting of $n$-qubits and replaces the lost qubit with the maximally mixed state, with probability $t$ between any two successive oracle invocations. 
This noise model is very drastic and, in some implementations, quite a pessimistic noise model, as it acts strongly correlated across the whole quantum register, somewhat similar to a noisy oracle.
It's important to note that quantum error correction techniques \cite{nielsen_quantum_2010, shor_fault-tolerant_1997} are not effective against partial depolarizing or dephasing noises, as these types of disturbances collectively impact the quantum computer. Such collective effects could be likened to a quantum system experiencing a sudden 'flash' of noise, possibly from fluctuating lasers or external magnetic field interference, which are plausible scenarios for quantum computers of moderate size. 

\subsection{State evolution with error}\label{subsec:state_evolve_noisy_grover}
At every step of the process, there is sequential application of the Grover operator $\mathcal{G}$ and subsequently, the depolarising channel error $\mathcal{D}_t$. The depolarising channel error follows the commutative property with any unitary operation while acting on the system \cite{cohn_grovers_2016}. Thus, in the context of Grover's algorithm subjected to noise, the resultant quantum state after the application of noise can be expressed as \( (\mathcal{D}_t\mathcal{G})^k(\rho) \), 
where \( \mathcal{G}(\rho) \) is defined by the expression \( \mathcal{G}(\rho) = ((\mathbb{I}_d - 2|\psi\rangle\langle\psi|)O_{\tau}) \rho ((\mathbb{I}_d - 2|\psi\rangle\langle\psi|)O_{\tau})^\dagger \). Considering this, 
Thus, After '$k$' iterations, the state is represented as:
\begin{equation}
\rho(k, t) = t^k \rho(k) + \left(1 - t^k\right) \frac{I}{d},
\end{equation}
where \(\rho(k)\) is as defined in eq.(\ref{eqn: ideal_grover_state_k}).
\subsection{Success probability}
Consequently, the probability of successfully measuring the target index of the oracle, denoted as \( \tau \), is given by the formula:
\begin{equation}\label{probability(k, (1-t), d)_noisy}
    P(k, (1-t), d) = \frac{1}{d} \sum_{x=1}^{d} \langle \tau | (\mathcal{D}_t\mathcal{G})^k(|\psi\rangle\langle\psi|) |\tau\rangle.
\end{equation}
The above expression denotes the success probability of Grover's search algorithm running on an input space $\rho$ of dimension $d=2^n$ after $k$ iterations, where \( d \) represents the total number of possible oracles. Here, we use $(1-t)$ to describe the depolarising noise strength for the duration of the algorithm where
$t$ is the channel parameter of the Total partial Depolarising Channel used to describe error \textit{at each step}, as given in \ref{eqn:noisy_grover_with_TPDCh}). Using $(1-t)$ as a parameter in $P(k,(1-t),d)$ will make the subsequent analysis of the effect of noise on probability more intuitive because we will be able to analyze and plot functions of $1-t)$ (error probability/ noise rate) instead of functions of $t$. This can be seen in the plots for framework 1 (fig.\ref{fig: F1_xi}) and framework 2 (fig.\ref{fig: F2_omega_Pvs(1-t)})\\
This formulation \ref{probability(k, (1-t), d)_noisy} averages the success probability over all \( d \) oracles, assuming that each oracle is equally likely a priori. This approach, focusing on the average success probability across all oracles, contrasts with alternative methodologies considering the minimal success probability, which seeks to minimize all oracles. The selection of the average success metric aligns with the perspective adopted in references such as \cite{zalka_grovers_1999}, differing from the minimal success probability approach seen in  \cite{boyer_tight_1998, nielsen_quantum_2010}.
\begin{eqnarray}
&&P(1,(1-t),d){}\nonumber 
= \frac{1}{d}\sum_{x=1}^d \langle x|(\mathcal{D}_t \mathcal{G}_x) (\rho)|x\rangle{}\nonumber\\&&
= t\cdot p(1,0,d) + (1-t)\frac{1}{d}
\qquad \text{[}\because\;p(1,0,d)\text{ is same as \ref{ideal_grover_success_probability}]}\nonumber\\&&
= t\cdot \sin\left((2(1)+1)\arcsin\left(\frac{1}{\sqrt{d}}\right)\right) + (1-t)\frac{1}{d}.
\end{eqnarray}

 For the above noise model, when considering the evolution process \( (\mathcal{D}_t\mathcal{G}_x)^k \) \ref{subsec:state_evolve_noisy_grover}, it can be expanded into a summation encompassing $2^k$ distinct histories. Given that every individual term in this summation contributes positively, a lower bound can be ascertained by retaining the term consisting of noise-free components \(t \rho\) in each product segment. This leads us to the inequality for the success probability \( P(k, (1-t), d) \) as follows:
\begin{equation}
 P(k, (1-t), d) \geq t^k \sin^2 \left( (2k + 1) \arcsin \frac{1}{\sqrt{d}} \right). 
 \end{equation}
When examining partial depolarizing noise, the probability of success can be precisely calculated \cite{vrana_fault-ignorant_2014}. We consider the $2^k - 1$ terms not included, expressed as \( \frac{1}{d} \sum_{x=1}^{d} (1-t)^m (t)^{k-m} \langle x | \mathbb{I}_d^{d} | x \rangle \), which simplifies to \( \frac{1}{d} (1-t)^m(t)^{k-m} \). Here, m denotes the instances of noise interference out of k iterations for a particular term. These specific terms are indicative of scenarios where noise introduces the maximally mixed state at certain junctures, and this holds given that both $\mathcal{D}_t$ and $\mathcal{G}_x$ \cite{vrana_fault-ignorant_2014} are unital. Considering the aggregate of the coefficients of these terms, which amounts to \( 1 - t^k \), we can deduce the precise probability of success for this noise model as,
\begin{equation}
    P(k, (1-t), d) = (1 - t^k) \frac{1}{d} + t^k \sin^2 \left( (2k + 1) \arcsin \frac{1}{\sqrt{d}} \right).
\end{equation}




\section{Application of Quantum Switch to Grover Search Algorithm} \label{sec:application_q_switch_grovers}
In this section, we develop two theoretical frameworks to explore and demonstrate the potential of quantum switches to mitigate noise accumulating in iterations of the Grover search algorithm. 
For this, we assume the resultant noise at every iteration to be originating in discrete steps within the iteration such that it can be modeled as a composition of two depolarising channels.
There are multiple steps in Grover's Search iteration/ subroutine. To control the noise coherently we assume that we can put the multiple steps in a Grover iteration into superposition. As discussed previously \ref{eqn: ideal_grover_state_k}, a Grover iteration consists of at least two discrete steps or subroutines For this, we assume the resultant noise at every iteration to be originating in discrete steps within the iteration such that it can be modeled as a composition of two depolarising channels.
\noindent The differences between the two frameworks become relevant only at the end of the first Grover iteration, where we first get the following choice: either (1) making a measurement and postselection now to trace out the switch correlated with the input state so that input to the next iteration is just the input state after first Grover iteration or (2) postponing the measurement and postselection till the end of the algorithm, such that the input to the next iteration will be the joint state of the input state correlated with the control qubit. This distinction will become clear as we further this discussion, but before delving into the specifics of each framework, we will establish and elaborate on certain assumptions common to both frameworks.\\

\noindent \textbf{First Iteration: (k=1):} Let us look at the first iteration of Grover's search algorithm running on an unstructured database. The input system will be an $n$-qubit system, represented by a $d$ dimensional qudit system where $d = 2^n$. This is because a one-to-one mapping exists between the $n$-qubit system and $d=2^n$ level qudit system. We model the noise in the iteration as a composition of two total depolarizing channel errors $(1-\sqrt{t})$. 
Thus, if we consider $t$ to be the channel parameter for the depolarizing channel, we can model noise in the first iteration as a channel, 
\begin{equation}\
    \mathcal{D}_{t}(\rho) = t \rho + (1 - t) \mathrm{Tr}[\rho] \dfrac{\mathbb{I}_d}{d} = \mathcal{D}_{\sqrt{t}}(\mathcal{D}_{\sqrt{t}}(\rho)) \label{depolarising_noise}.
\end{equation} 
\noindent This resultant noise could be assumed to be a combination of two identical depolarizing channel noises, each with channel parameter $\sqrt{t}$, error probability $1-\sqrt{t}$. These noises can be expressed as the channel, 
\begin{equation}\label{eqn:decomposed_error}
    \mathcal{D}_{\sqrt{t}}(\rho) = \sqrt{t} \rho + (1 - \sqrt{t}) \mathrm{Tr}[\rho] \dfrac{\mathbb{I}_d}{d} = \sum\limits_{i = 0}^{d^2} K_i \rho K_i^\dagger. 
\end{equation} 
Here the set of Kraus operators, $\{K_i\}$, for the above channel will be, $K_0 = t^{\frac{1}{4}}\mathbb{I}_d$ and $K_i = \sqrt{\dfrac{(1-\sqrt{t})}{d^2}}U_i$ \, where \{$i = 1,2,\ldots d^2$\}. Thus, we can express the Kraus operators of the switch $\mathcal{S}$ as,
\begin{eqnarray}
  \mathcal{W}_{ij} 
  &&= K_i K_j \otimes \ket{0}\bra{0} + K_j K_i \otimes \ket{1}\bra{1}{}\nonumber\\
  &&= \dfrac{(1-\sqrt{t})}{d^2}\Bigg (U_i U_j \otimes \ket{0}\bra{0} + U_j U_i \otimes \ket{1}\bra{1}\Bigg), 
\end{eqnarray}
\noindent for  $i,j =  0,1,2\ldots d^2$. Here, $\{K_i\}$ and $\{K_j\}$ are Kraus operators for the two identical channels given in (\ref{eqn:decomposed_error}). Suppose that the control system is fixed to the state, $\rho_c = \ket{c}\bra{c}$, where, $\ket{c}=\sqrt{\theta}\ket{0} + \sqrt{\overline{\theta}}\ket{1}$. We prepare the $n$-qubit system, which is input to Grover's search, in the state $\rho(0)$, which can be considered a $d$ dimensional qudit system. 
After the first Grover iteration, the system's state is supposed to be $G\rho(0)G^\dagger = \rho(1)$ as eq.(\ref{eqn: ideal_grover_state_k}. But, If we consider the noise as in eq.(\ref{depolarising_noise}), the resultant state can be denoted by, 
\begin{equation}\label{eqn:noise_decomposition}
    D_{t}(\rho(1)) = t \rho(1) + (1 - t) \mathrm{Tr}[\rho(1)] \dfrac{\mathbb{I}_d}{d} = \mathcal{D}_{\sqrt{t}}(\mathcal{D}_{\sqrt{t}}(\rho(1))). 
\end{equation}    
Now, If we consider a quantum switch applied on these two decomposed channels, the resultant state  $\mathcal{S} (\mathcal{D}_{\sqrt{t}}, \mathcal{D}_{\sqrt{t}})(\rho(1) \otimes \rho_{c})$ is given by,

\begin{widetext}
\begin{eqnarray}
    &&\mathcal{S} (\mathcal{D}_{\sqrt{t}}, \mathcal{D}_{\sqrt{t}})(\rho(1) \otimes \rho_{c}) 
    = \mathlarger{\mathlarger{\sum\limits}}_{i, \,j = 0}^{d^2} \mathcal{W}_{ij}(\rho(1) \otimes \rho_{c}) \mathcal{W}_{ij}^{\dagger}\\
    &&= \mathlarger{\mathlarger{\sum\limits}}_{i, j = 0}^{d^2} \Bigl\{ \bigl(K_i K_j\otimes \ket{0} \bra{0} + K_j K_i \otimes \ket{1} \bra{1}\bigl)\bigl(\rho(1) \otimes \rho_{c}\bigl)\bigl(K_i K_j\otimes \ket{0} \bra{0} + K_j K_i \otimes \ket{1} \bra{1}\bigl)^{\dagger}\Bigl\}
\end{eqnarray}
\begin{eqnarray}\label{fourier_measured_1}\label{eqn: init_switched_1}
    \mathcal{S}(\mathcal{D}_{\sqrt{t}},\mathcal{D}_{\sqrt{t}})(\rho(1) \otimes \rho_{c}) &&= (1-\sqrt{t})^2\left(\mathrm{Tr}[\rho(1)]\frac{\mathbb{I}_d}{d}\otimes
    \left(\theta\ket{0}\bra{0}+\overline{\theta}\ket{1}\bra{1}\right)
    \nonumber\right.\\
    &&\quad\left.
    +\frac{\rho(1)}{d^2}\otimes\sqrt{\theta\overline{\theta}}\left(\ket{0}\bra{1}+\ket{1}\bra{0}\right)\right)\nonumber\\
    &&\quad+2\sqrt{t}(1-\sqrt{t})\left(\mathrm{Tr}[\rho(1)]\frac{\mathbb{I}_d}{d}\otimes\rho_c\right)+t(\rho(1)\otimes\rho_c).    
\end{eqnarray}
\end{widetext}

\noindent For more detailed calculations, refer to Appendix \ref{appendix:q_switch_k=1_calc}. The quantum switch facilitates the transfer of information not only within the input system $\rho$ or the control system $\rho_c$ but primarily through the correlations established between the output system and the control. These quantum correlations are crucial for the successful transfer of information. If the control qubit experiences decoherence within the computational basis $\{\ket{0}, \ket{1}\}$, the information becomes entirely inaccessible. However, for our study, we assume the control system will be noiseless for the duration of our investigation. Finally, we measure the control qubit in the basis $\{\ket{+}, \ket{-}\}$, and these measurements yield conditional states.

\begin{figure*}[htp]
\noindent \begin{mdframed}
[
        linecolor=black,linewidth=0.5pt, frametitlerule=true,
        apptotikzsetting={\tikzset{mdfframetitlebackground/.append style={
            shade,left color=white, right color=gray!20}}}, 
        frametitlerulecolor=black,
        frametitlerulewidth=0.5pt, innertopmargin=\topskip,
        frametitle={Framework 1, $F_\xi$},
]
\includegraphics[scale=0.65]{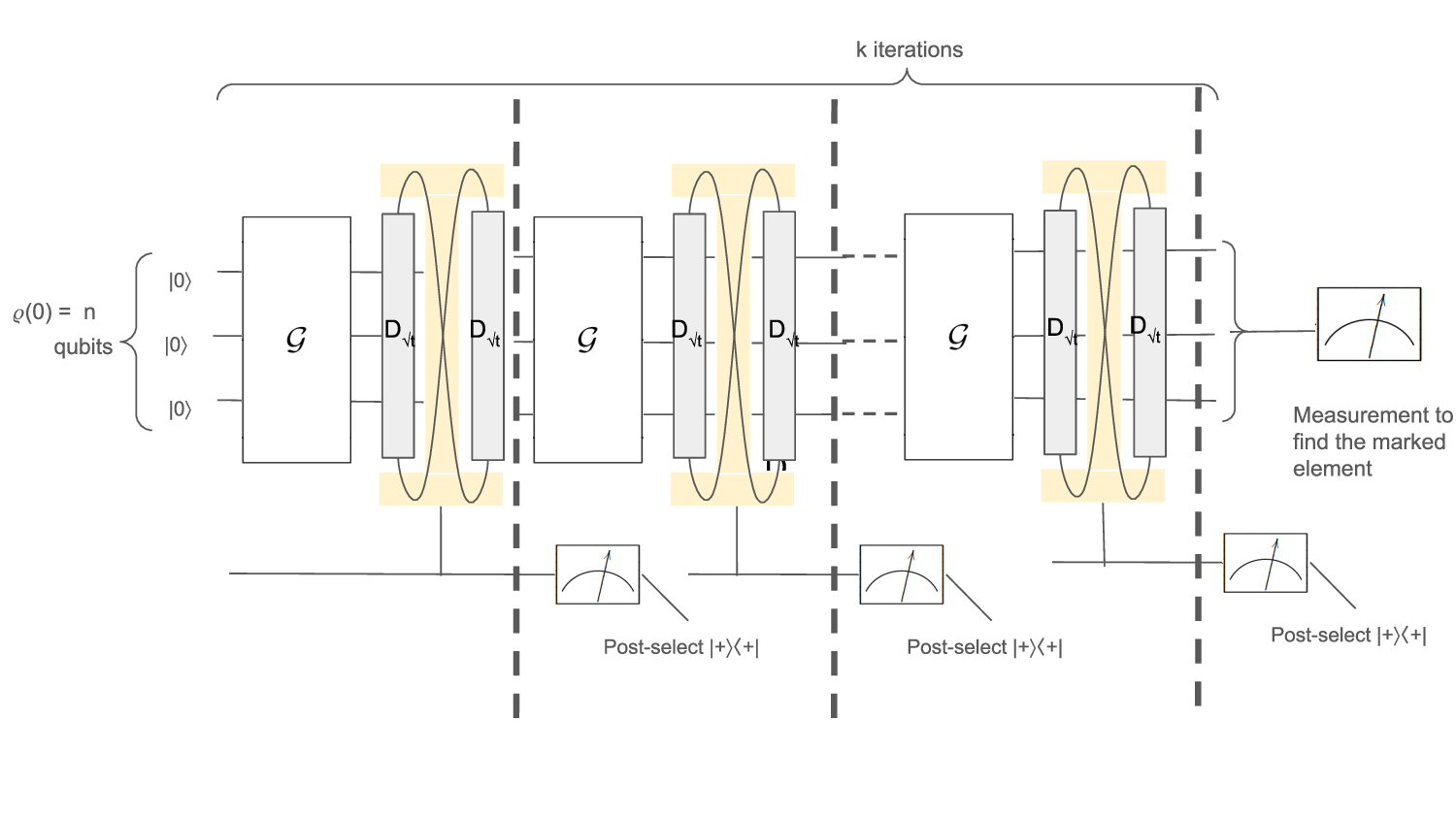}
\caption{\label{fig: noisy_grovers_framework1_every_iter} The figure depicts the application of switch to Grover's search algorithm (modeled by the Grover operator $\mathcal{G}$) to mitigate the noise that arises because of the total depolarizing channel. The channel is the ash-colored bar, while the yellow region represents the switch. The dotted line indicates that the post-selection happens at every step, which is specific to this framework. }
\end{mdframed}
\end{figure*}

\subsection{Framework 1, \texorpdfstring{$F_\xi$}{}: Measurement after every iteration }\label{subsec: F1_xi}
In the schematic diagram fig.(\ref{fig: noisy_grovers_framework1_every_iter}), we show the action of the Grover operator, along with switched noisy total depolarising channels in the first framework. Notice the measurement and postselection after each iteration.
\begin{figure*}[htp!]
\noindent \begin{mdframed}
[
        linecolor=black,linewidth=0.5pt, frametitlerule=true,
        apptotikzsetting={\tikzset{mdfframetitlebackground/.append style={
            shade,left color=white, right color=gray!20}}}, 
        frametitlerulecolor=black,
        frametitlerulewidth=0.5pt, innertopmargin=\topskip,
        frametitle={Framework-1: Comparison of Success probability between the switch and non-switch scenario},
]
    \includegraphics[scale=0.9]{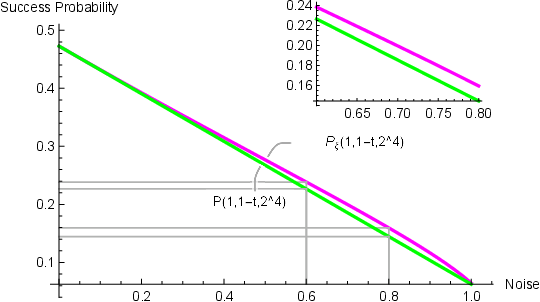}
    \includegraphics[scale=0.9]{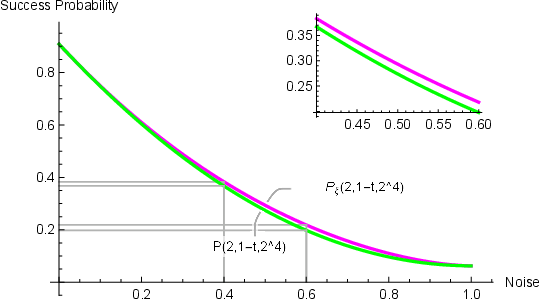}
    \includegraphics[scale=0.9]{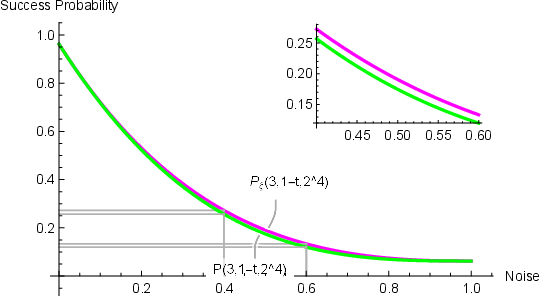}
    \begin{minipage}[b]{0.5\textwidth}
        \caption{ These plots show the effect of noise strength $(1-t)$ on the success probability of finding the target element in the search space in Noisy Grover's search algorithm. We are taking the search space to be $d = 2^4$, and thus the algorithm should stop at $k_Gr = \frac{\pi}{4}\sqrt{16} = \pi \approx 3$ iterations. The Plots from left to right show these variations for different iterations: k = 1 (top-left), k = 2 (top-right), and 3 (bottom-left). Here, the green curve represents the Success probability without using any switches, and the magenta curve represents the success probability on applying switches as described in fig.\ref{fig: noisy_grovers_framework1_every_iter}}
        \label{fig: F1_xi}
    \end{minipage}
    \end{mdframed}
\end{figure*}
\subsubsection{State Evolution with Error}


\noindent In this framework, we measure the basis $\{\ket{+}, \ket{-}\}$ at the end of every iteration. The state of the system at the end of the first iteration is given by,
\begin{eqnarray}\label{eqn:switched1_noisy_grover_evolve_M}
    \rho_{\xi}(1)^\pm=\frac{\bra{\pm}\mathcal{S}(D_{\sqrt{t}}, D_{\sqrt{t}})(\rho(1)\otimes\rho_{c})\ket{\pm}}{\mathrm{Tr}\left[\bra{\pm}\mathcal{S}(D_{\sqrt{t}}, D_{\sqrt{t}})(\rho(1)\otimes\rho_{c})\ket{\pm}\right]}.
\end{eqnarray}

\noindent Here, $\xi$ in the subscript denotes the stepwise switched channel framework. 
%
%
\noindent If we substitute $t$ as $0$, in eq.(\ref{fourier_measured_1}), it yields the same results as \cite{ebler_enhanced_2018}, which uses completely depolarizing channels for demonstrating enhancement in communication. 
However, in the case of Grover's search algorithm, we will consider t as a variable noise parameter for our analysis. The density matrix after measurement in the basis $\{\ket{+}, \ket{-}\}$ is given by eq.(\ref{eqn:switched1_noisy_grover_evolve_M}),
\begin{equation}
\rho_\xi(1)^{(\pm)}
= N/M,
\end{equation}
where, 
\begin{eqnarray*}
    &&N=\frac{1}{2}\Biggl\{\left(t \pm2\sqrt{\theta\overline{\theta}}\left(\left(\frac{1-\sqrt{t}}{d}\right)^2+t\right)\right)\rho(1)\nonumber\\
    &&\quad+\left((1-t) \pm2\sqrt{\theta\overline{\theta}} 2\sqrt{t}(1-\sqrt{t})\right)\mathrm{Tr}[\rho(1)]\frac{\mathbb{I}_d}{d}\Biggl\}
\end{eqnarray*}
and 
\begin{eqnarray*}
    &&M=\frac{1}{2}\Biggl\{\left(t \pm2\sqrt{\theta\overline{\theta}}\left(\left(\frac{1-\sqrt{t}}{d}\right)^2+t\right)\right)\nonumber\\
    &&\qquad+\left((1-t) \pm2\sqrt{\theta\overline{\theta}} 2\sqrt{t}(1-\sqrt{t})\right)\Biggl\}
\end{eqnarray*}
Now we have, after post selecting on +

\begin{equation}
\rho_\xi(1)^{+}= f_\xi(t)\rho(1)+ (1-f_\xi(t))\frac{\mathbb{I}_d}{d}.
\end{equation}   
where, 
\begin{equation}
    f_\xi(t)= \frac{\left(\frac{1-\sqrt{t}}{d}\right)^2 + 2t}{\left(\frac{1+(t-2\sqrt{t})(1-d^2)}{d^2}+1\right)}.
\end{equation}


\subsubsection{Success Probability}

\noindent For this switched channel framework, 
we have the success probability $P_\xi$ for Grover's search algorithm assisted by this switched channel framework as, 
\begin{eqnarray}
    P_\xi(1,(1-t),d)= f_\xi(t)
    P(1,0,d)\nonumber
    + \frac{\left(1- f_\xi(t)
    \right)}{d}.
\end{eqnarray}

\noindent \textbf{Generalisation to $k$ iterations: } Previously, we have shown how the framework behaves when there is only 1 iteration. We extend this to $k$ iterations. In other words, if we do postselection measurement at the $k^{th}$ step, then the final success probability is given by,
\begin{eqnarray}
    P_\xi(k,(1-t),d) = 
    \left(f_\xi(t)\right)^k
    P(k,0,d)\nonumber+\frac{1- \left(f_\xi(t)\right)^k}
    {d}. \quad 
\end{eqnarray}

\noindent Here, in this framework, we take a measurement (without disturbing the output state after every iteration) and then do a postselection, which is a resource-heavy operation. This, in turn, also destroys the correlation between the switch and the state at every step, and we can keep resetting the same quantum state as a switch repeatedly until the end of the algorithm. So, we will be proposing a better framework in the next subsection.\\
\noindent We plot the fig.(\ref{fig: F1_xi}) comparing the success probability of the switched model of framework 1 and the success probability of the noisy Grover iteration when there is no switch (we take $d=2^4$ as an example. The switch gives some advantages in restoring the probability of success for the first iteration. However, the noise gradually reduces the advantage as the algorithm undergoes further iterations.

\subsection{Framework 2, \texorpdfstring{$F_\omega$}{}: Measurement at the end }\label{subsec: F2_omega}
\begin{figure*}[htp!]
\noindent \begin{mdframed}
[
        linecolor=black,linewidth=0.5pt, frametitlerule=true,
        apptotikzsetting={\tikzset{mdfframetitlebackground/.append style={
            shade,left color=white, right color=gray!20}}}, 
        frametitlerulecolor=black,
        frametitlerulewidth=0.5pt, innertopmargin=\topskip,
        frametitle={Framework-2, $F_\omega$},
]
\includegraphics[scale=0.65]{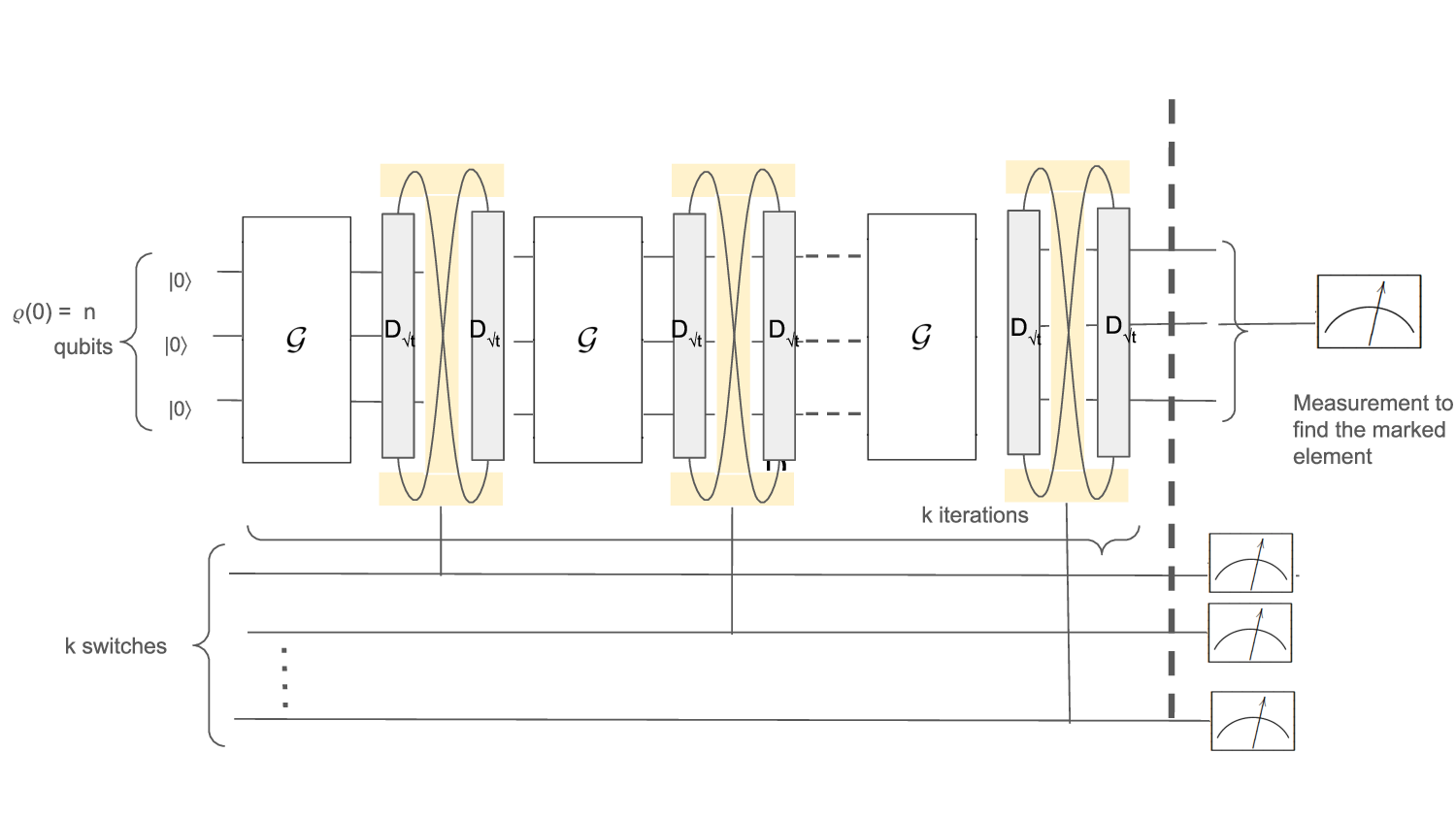}
\caption{\label{fig: noisy_grovers_framework2_last_iter} The figure depicts the application of the switch to Grover's search algorithm (modeled by the Grover operator $\mathcal{G}$). The channel is the ash-colored bar, while the yellow region represents the switch. The dotted line at the end indicates that the postselection happens at the very end, which is specific to this particular framework. Here, we take a register of $k$ switches, thus allowing us to preserve the system and switch correlation until the end.}
\end{mdframed}
\end{figure*}

In the schematic diagram fig.(\ref{fig: noisy_grovers_framework2_last_iter}), we show the action of the Grover operator, along with switched noisy total depolarising channels in the second framework. Notice the measurement and post-selection at the end of the algorithm on the complete joint state combining the input state with $k$ quantum switches, where $k$ is the number of Grover iterations the state has traversed through.
\noindent Our previous framework for analyzing the system after each Grover's search iteration involved performing measurements and postselections at every step. So, the gain in the probability of success may be attributed to the state's post-selection. However, we have now adopted a different approach for the second framework to circumvent this. In this new framework that we propose in this section, we see the cumulative effect of $k$ switches after $k$ iterations by measuring them on the Hadamard basis at the very end. Instead of measuring and post-selecting the control qubit after each iteration, we use multiple quantum switches and hold off on these actions until the end of all iterations. By doing so, the system can undergo a sequence of Grover search iterations without intermediate collapse, which may result in dynamics different from those of the first framework.

\subsubsection{Analysing the output state obtained after the first iteration}
\noindent Before we go into the details of this framework, we start with analyzing the output obtained after the first iteration, which will be common in both. The result above demonstrates a dependence on the parameter $\theta$ of the control qubit \(\lvert c \rangle\), solely through the coherent indefiniteness \cite{frey_indefinite_2019} denoted as \(\sqrt{\theta \overline{\theta}}\) in eq.(\ref{fourier_measured_1}). It is apparent that the optimal choice for the maximum switch setting, i,e 2\(\sqrt{\theta \overline{\theta}} = 1\), proves to be advantageous across all values of $t$ and dimensions $d$.
\noindent Notably, any positive degree of indefiniteness (\(\theta > 0\)) proves advantageous, signifying that increased indefiniteness yields enhanced benefits. Moreover, it becomes evident that maximum indefiniteness (\(\theta = \frac{1}{2}\), 2\(\sqrt{\theta \overline{\theta}} = 1\)) provides the most favorable conditions for this purpose. 

Taking the $|+\rangle$ component of the measurement. Post-selecting and correcting based on the $+$ or $-$ part of the measurement.
 We should also note that even post-selecting on - would have led to similar results with an inverted phase and thus we can consider the further results without loss of generality.
This output $\mathcal{S}(\mathcal{D}_{\sqrt{t}}, \mathcal{D}_{\sqrt{t}})(\rho(1) \otimes \rho_c)$ in eq.(\ref{eqn: init_switched_1}) can be rearranged as:
\begin{eqnarray}\label{rearranged_eqn: init_switched_1}
    &&\mathcal{S}(\mathcal{D}_{\sqrt{t}},\mathcal{D}_{\sqrt{t}})(\rho(1) \otimes \rho_{c})\nonumber\\
    &&= \left\{t\rho(1) + (1-t)\mathrm{Tr}[\rho(1)]\frac{\mathbb{I}_d}{d}\right\}\nonumber \otimes\left(\theta\ket{0}\bra{0} + \overline{\theta}\ket{1}\bra{1}\right)\nonumber\\
    &&+\left\{\left(\left(\frac{1-\sqrt{t}}{d}\right)^2+t\right)\rho(1) + 2\sqrt{t}(1-\sqrt{t})\mathrm{Tr}[\rho(1)]\frac{\mathbb{I}_d}{d}\right\}\nonumber\\ &&\otimes\sqrt{\theta\overline{\theta}}\left(\ket{0}\bra{1}+\ket{1}\bra{0} \right).
\end{eqnarray}
The term \( \bigl(1-t\bigl)\mathrm{Tr}[\rho]\dfrac{\mathbb{I}_d}{d}+t\rho(1) \) in eq.(\ref{fourier_measured_1}) represents the output after first iteration corresponding to the parameter \(\theta\) under the scenario where the channel order remains entirely definite (\(\theta\) = 0 or 1). Meanwhile, the second term in eq.(\ref{fourier_measured_1}) signifies the additional information gained due to any degree of ICO ($0 <\theta< 1$). These observations emphasize that within the context of Grover's search algorithm, the quantum switch functions as a mitigator of noise by facilitating more information transfer.

\subsubsection{Block Matrix Operations and Notations}\label{subsubsec: notation}

We introduce a set of notations that enable a more concise representation of states within the framework of switched quantum channels that capture the recursive operations on quantum states. This formalism is based on the structure of block matrices and the operations performed on them, which are pivotal for understanding the dynamics of quantum states under the influence of noise and control mechanisms.\\ 
\begin{enumerate}
    \item Consider a block matrix \(A\), composed of \(2^k \times 2^k\) blocks, with each block \(A_{ij}\) being a \(d \times d\) matrix. This structure enables a fine-grained representation of quantum operations on multipartite states. To facilitate the analysis of such operations, we define the element-wise trace operation, \( \mathrm{Tr}_{d \times d}(A) \), which transforms \(A\) into a new \( 2^k \times 2^k \) matrix \(T\). Each element \(T_{ij}\) of \(T\) is obtained by taking the trace of the corresponding \(d \times d\) block \(A_{ij}\) in \(A\). Formally, this operation is expressed as:
    \begin{multline}
        \mathrm{Tr}_{d \times d}(A) = T \\\text{where } T_{ij} = \mathrm{Tr}[A_{ij}]\; \forall \; i \text{ in } 1, 2, \ldots, M \text{ and } j \text{ in } 1, 2, \ldots, N.
    \end{multline}
    
    \item We introduce notations for functions of $t$ to compactly describe the effects of certain quantum operations on states. \(r_{\rho}\) and \(r_{\mathbb{I}}\) are defined as:
    \begin{equation}\label{def: f(t)r(t)_notation}
    r_{\rho} = \left(\dfrac{1-\sqrt{t}}{d}\right)^2 + t, \quad r_{\mathbb{I}} = 2\sqrt{t}(1-\sqrt{t}),    
    \end{equation}
    where \(\sqrt{t}\) is the noise parameter related to the depolarising channel $\mathcal{D}_{\sqrt{t}}$. 
    
    \item Similarly, we define \(f_{\rho}\) and \(f_{\mathbb{I}}\) to maintain symmetry in our notations:
    \begin{equation}
    f_{\rho} = t, \quad f_{\mathbb{I}} = (1-t).    
    \end{equation}
    
    \item Building upon these definitions, we define and examine the action of a fictitious operation \(\mathcal{F}\) on the joint quantum state \(\rho_{\omega}\) consisting of the original $d$-dimensional input state correlated with the $k$ switches, which is represented as:
    \begin{equation}
    \mathcal{F}(\rho_{\omega}) = f_{\mathbb{I}}\mathrm{Tr}_{d \times d}[\rho_{\omega}]\dfrac{\mathbb{I}_d}{d}+f_{\rho} \rho_{\omega},
    \end{equation} 
    \noindent demonstrating how the operation mixes the state with its trace over the identity matrix. 
    
    \item Likewise, another operation \(\mathcal{R}\) is characterized by:
    \begin{equation}\label{def: F(rho)R(rho)_notation}
    \mathcal{R}(\rho_{\omega}) = r_{\rho}\rho_{\omega}+r_{\mathbb{I}}\mathrm{Tr}_{d \times d}[\rho_{\omega}]\dfrac{\mathbb{I}_d}{d},    
    \end{equation}

    \item Thus, we can represent the expression for $\mathcal{S}(\mathcal{D}_{\sqrt{t}},\mathcal{D}_{\sqrt{t}})(\rho(1) \otimes \rho_{c})$ \(\rho(1) \otimes \rho_{c}\), given in eq.(\ref{rearranged_eqn: init_switched_1}), in a block matrix form to illustrate the resulting entangled state. The operation is written as:
    \begin{equation}\label{eqn: k=1_result}
    \mathcal{S}(\mathcal{D}_{\sqrt{t}},\mathcal{D}_{\sqrt{t}})(\rho(1) \otimes \rho_{c}) = \begin{bmatrix}
        \theta \mathcal{F}(\rho(1))
        & \sqrt{\theta\overline{\theta}}\mathcal{R}(\rho(1))\\
        \sqrt{\theta\overline{\theta}}\mathcal{R}(\rho(1)) 
        & \overline{\theta}\mathcal{F}(\rho(1))
        \end{bmatrix},    
    \end{equation}
    where \(\theta\) and \(\overline{\theta} = 1-\theta\) denote the coefficients that modulate the effect of operations \(\mathcal{F}\) and \(\mathcal{R}\) on the state. 
    
    In further text, we will be using shorthand to write this output as:
    \begin{equation}
       \rho_{\omega,  1}(1) = \mathcal{S}(\mathcal{D}_{\sqrt{t}},\mathcal{D}_{\sqrt{t}})(\rho(1) \otimes \rho_{c}) \label{def: rho_omega1(1)}
    \end{equation}
    Here, the $\omega$ in subscript highlights the second framework where we make measurements at the end. The integer after that (1, in this case) denotes the number of switches correlated to the input state. The integer in the bracket denotes the number of Grover iterations applied on the search space, the input state. 
\end{enumerate}
This representation not only captures the interaction between different components of the state but also helps in verifying that the output is a valid density matrix with unit trace, adhering to the principles of quantum mechanics.

\subsubsection{State Evolution with Error}

Here, we use a register containing $k$ switches, and it should be noted that this framework is identical to the previous framework for the first iteration because we're measuring the control state after the first iteration in the previous framework. The stark departure from the previous framework can be noticed if we look at further iterations where we will consider the quantum state consisting of the output after the first iteration ($\rho_1(1)$) and the control state ($\rho_{c_1}$) as the input to the next noisy Grover iteration. This contrasts with the previous framework, where we measure the control state and perform postselection, thus destroying the correlation between the input state and the quantum switch. We are only able to calculate the states obtained on running the Grover's Search Algorithm in this framework till $k=3$ as the matrix calculations blow up exponentially with the number of iterations. \\ 

\noindent\textbf{Second Iteration ($k = 2$): }We know the action of the switched noisy channels from the first iteration. This output, where the state representing the search space is now correlated with the first switch $\rho_{c_1}$, will be the input to the next iteration. Thus, unlike the first framework where in the next iteration, $\mathcal{G}$ is applied again on the search space $\rho_{\xi}(1)$, after the switch is traced out, in this framework, the next Grover iteration will act as   $\Bigl(\mathcal{G}\otimes\mathbb{I}_d\Bigl)$ on $\rho_{\omega, 1}(1)$ to make sure only the search space goes through the Grover iteration ($\mathcal{G}$). Following the notation defined in \ref{def: rho_omega1(1)}, We denote this output as $\rho_{\omega, 1}(2)$. After this, the noise acting on this iteration will be a combination of ($\mathcal{D}_{\sqrt{t}}\otimes\mathbb{I}_d$), as discussed earlier \ref{eqn:noise_decomposition}.\\ 
Here, We assume the quantum switches are noiseless for the time scale we're considering. Now, we use another switch $\rho_{c_2}$ to put these noisy channels in superposition as We apply the noise, $\bigl(D_{\sqrt{t}} \otimes \mathbb{I}_d\bigl)$ because the switch is noiseless. We can express the Kraus operators of the switch with two identical noise $\bigl(D_{\sqrt{t}} \otimes \mathbb{I}_d\bigl)$.

\begin{eqnarray}
\mathcal{W}_{ij}^{(2)} &&= {}\nonumber
\Bigl(K_i \otimes \mathbb{I}_d\Bigl)\Bigl(K_j \otimes \mathbb{I}_d\Bigl) \otimes \ket{0}\bra{0}\nonumber\\
&&\quad+ \Bigl(K_j \otimes \mathbb{I}_d\Bigl)\Bigl(K_i \otimes \mathbb{I}_d\Bigl) \otimes \ket{1}\bra{1}{}\\ &&=\Bigl(K_iK_j\otimes\mathbb{I}_d\Bigl)\otimes\ket{0}\bra{0}+\Bigl(K_jK_i\otimes\mathbb{I}_d\Bigl)\otimes\ket{1}\bra{1}\nonumber\\
\end{eqnarray}
\begin{eqnarray}  
\rho_{\omega, 2}(2)
&&=\mathcal{S}\left(\mathcal{D}_{\sqrt{t}}\otimes\mathbb{I}_d, \mathcal{D}_{\sqrt{t}}\otimes\mathbb{I}_d\right)(\rho_{\omega, 1}(2) \otimes \rho_{c_2}){}\\
&&= \mathlarger{\sum\limits}_{i, j =0}^{d^2}\mathcal{W}_{i j}^{(2)}\Bigl(\rho_{\omega, 1}(2) \otimes \rho_{c}\Bigl)\mathcal{W}_{i j}^{\dagger}{}
\end{eqnarray}

\noindent We know the action of the switched noisy channels from the k = 1 iteration if we apply the same action recursively; for k = 2, we get this. 

\begin{widetext}
\begin{eqnarray}
\rho_{\omega, 2}(2) =&& 
    \left\{\underbrace{f_{\rho} \begin{bmatrix} \theta \mathcal{F}(\rho(2)) & \sqrt{\theta\overline{\theta}}\mathcal{R}(\rho(2))\\ \sqrt{\theta\overline{\theta}}\mathcal{R}(\rho(2)) & \overline{\theta}\mathcal{F}(\rho(2))\end{bmatrix}}
    _{f_{\rho}(.) + } 
    + \underbrace{f_{\mathbb{I}} \frac{\mathbb{I}_d}{d}\otimes\begin{bmatrix}\theta \mathrm{Tr}[\mathcal{F}(\rho(2))] & \sqrt{\theta\overline{\theta}} \mathrm{Tr}[\mathcal{R}(\rho(2))]\\\sqrt{\theta\overline{\theta}} \mathrm{Tr}[\mathcal{R}(\rho(2))] & \overline{\theta} \mathrm{Tr}[\mathcal{F}(\rho(2))]\end{bmatrix}}
    _{f_{\mathbb{I}} \frac{\mathbb{I}_d}{d} \otimes \mathrm{Tr}_{d\times d}[.]} \right\}\nonumber\\
&&\otimes \left(\theta\ket{0}\bra{0} + \overline{\theta}\ket{1}\bra{1}\right)\nonumber\\
    &&+\left\{\underbrace{r_{\rho} \begin{bmatrix}\theta \mathcal{F}(\rho(2)) & \sqrt{\theta\overline{\theta}}\mathcal{R}(\rho(2))\\ \sqrt{\theta\overline{\theta}}\mathcal{R}(\rho(2)) & \overline{\theta}\mathcal{F}(\rho(2)) \end{bmatrix}}
    _{r_{\rho}(.) + } 
    + \underbrace{r_{\mathbb{I}} \frac{\mathbb{I}_d}{d}\otimes\begin{bmatrix}\theta \mathrm{Tr}[\mathcal{F}(\rho(2))] & \sqrt{\theta\overline{\theta}} \mathrm{Tr}[\mathcal{R}(\rho(2))]\\ \sqrt{\theta\overline{\theta}} \mathrm{Tr}[\mathcal{R}(\rho(2))] & \overline{\theta} \mathrm{Tr}[\mathcal{F}(\rho(2))] \end{bmatrix}}
    _{r_{\mathbb{I}} \frac{\mathbb{I}_d}{d} \otimes \mathrm{Tr}_{d\times d}[.]}\right\}\nonumber\\
&&\otimes\sqrt{\theta\overline{\theta}}\left(\ket{0}\bra{1} + \ket{1}\bra{0}\right)\\
\rho_{\omega, 2}(2) =&& \left\{f_{\rho}\rho_{\omega, 1}(2)  + f_{\mathbb{I}} \frac{\mathbb{I}_d}{d}\otimes \mathrm{Tr}_{d\times d}[\rho_{\omega, 1}(2)] \right\}
 + \otimes \left(\theta\ket{0}\bra{0} + \overline{\theta}\ket{1}\bra{1}\right)\nonumber\\
&&+\left\{r_{\rho} \rho_{\omega, 1}(2) + r_{\mathbb{I}} \frac{\mathbb{I}_d}{d}
     \otimes \mathrm{Tr}_{d\times d}[\rho_{\omega, 1}(2)]\right\} 
\otimes\sqrt{\theta\overline{\theta}}\left(\ket{0}\bra{1} + \ket{1}\bra{0}\right)
\end{eqnarray}
\end{widetext}

\noindent See Appendix \ref{appendix: switched_kgt1}, for detailed calculation. 
We can use eqn.(\ref{def: F(rho)R(rho)_notation}) and expand the previous notation of $\mathcal{F}$ and $\mathcal{R}$ to write shorthand notation (similar to eqn.\ref{def: rho_omega1(1)}) for the above result as: 
\begin{equation} \label{eqn: k=2_result}
    \rho_{\omega, 2}(2) = \begin{bmatrix}
   \theta \mathcal{F}_{(2)}(\rho_{\omega, 1}(2)) & \sqrt{\theta\overline{\theta}} \mathcal{R}_{(2)}(\rho_{\omega, 1}(2)) \\ \sqrt{\theta\overline{\theta}} \mathcal{R}_{(2)}(\rho_{\omega, 1}(2)) & \theta \mathcal{F}_{(2)}(\rho_{\omega, 1}(2))
\end{bmatrix},    
\end{equation}
where the matrix in eq.(\ref{eqn: k=2_result}) is a $4d$-dimensional block matrix which is again analogous to matrix in eq.(\ref{eqn: k=1_result}). Now, if we want to stop the algorithm at this stage and analyze the state for success probability, we will need to make the measurement, assuming the quantity $2\sqrt{\theta\overline{\theta}} = 1$ for maximum indefiniteness as discussed before
%
\begin{eqnarray}
    \rho_{\omega}(2) =&& \frac{(\bra{+}^{\otimes2})\rho_{\omega, 2}(2)(\ket{+}^{\otimes2})}{\mathrm{Tr}\left[(\bra{+}^{\otimes2})\rho_{\omega, 2}(2)(\ket{+}^{\otimes2})\right]}\nonumber\\
    =&&  f_{\omega, 2}(t)\rho(2)
    + (1-  f_{\omega, 2}(t))\mathrm{Tr}[\rho]\frac{\mathbb{I}}{d},
\end{eqnarray}
    where, 
\begin{eqnarray}
    f_{\omega, 2}(t) =&& (f_{\rho}+r_{\rho})^2\Bigl((f_{\rho}+r_{\rho})^2+(f_\mathbb{I}+ r_\mathbb{I})\nonumber\\&&(1+r_{\rho}+ r_I) + f_{\rho} + r_{\rho})\Bigl)^{-1}
\end{eqnarray}

\noindent \textbf{Extending the output state to $k^{th}$ iteration:} The quantum operation applied to the output of $(k-1)^{th}$ (input to the $k^{th}$) iteration will be $\left(\mathbb{I}_d\otimes G D_t\right)\rho_{\omega, k-1}(k)\left(D_t^{\dagger}G^{\dagger}\otimes\mathbb{I}_d^{\dagger}\right)$.
If we apply that action recursively, (as the input of the $k^{th}$ iteration is the same as the output of the $(k-1)^{th}$ iteration) 

\noindent At the $k^{th}$ iteration, the state/system will be a $2^k*2^n = 2^{k+n}$ dimensional matrix. In terms of the number of terms, there will be $2^k*2^k = 2^{k+k} = 4^k$ terms. To study the accumulated error at the $k^{th}$ iteration, we will need to consider taking the state obtained after measurement and postselection of all the $k$ control states correlated with the output state after $k$ iterations. 
Here we can start unpacking the block matrices $\mathcal{F}$ and $\mathcal{R}$:
\begin{widetext}
\begin{equation}\label{eqn: F_k-sub}
    \mathcal{F}(\rho_{\omega, k-1}(k)) = f_{\rho}\underbrace{(\rho_{\omega, k-1}(k))}_{2^{k-1}d  -dimensional} + f_{\mathbb{I}} \underbrace{\frac{\mathbb{I}_d}{d}}_{d -dimensional} \otimes \underbrace{\mathrm{Tr}_{d \times d}[\rho_{\omega, k-1}(k)]}_{2^{k-1} -dimensional}.
\end{equation}
\begin{equation}\label{eqn: R_k-sub}
    \mathcal{R}(\rho_{\omega, k-1}(k)) = r_{\rho}\underbrace{(\rho_{\omega, k-1}(k))}_{2^{k-1}d  -dimensional} + r_{\mathbb{I}} \underbrace{\frac{\mathbb{I}_d}{d}}_{d -dimensional} \otimes \underbrace{\mathrm{Tr}_{d \times d}[\rho_{\omega, k-1}(k)]}_{2^{k-1} -dimensional}.
\end{equation}
\end{widetext}

We get a recursive relation using the above substitutions in (\ref{eqn: F_k-sub}) and (\ref{eqn: R_k-sub}). Again, for concise representation, we introduce a notation called block trace, which applies trace on all the $2 \times 2$ block matrices in the matrix. It leads to a matrix that is half the size of the input matrix, again keeping in line with the exponential nature of the calculations. We will reduce the matrix's dimension by half with each step. 

\begin{widetext}
\begin{eqnarray}
    &&(\bra{+}^{\otimes k}) \rho_{\omega, k}(k) (\ket{+}^{\otimes k}) = (\bra{+}^{\otimes k}) \begin{bmatrix}
     \theta \mathcal{F}(\rho_{\omega, k-1}(k)) & \sqrt{\theta\overline{\theta}} \mathcal{R}(\rho_{\omega, k-1}(k)) \\ \sqrt{\theta\overline{\theta}} \mathcal{R}(\rho_{\omega, k-1}(k)) & \overline{\theta} \mathcal{F}(\rho_{\omega, k-1}(k))
 \end{bmatrix} (\ket{+}^{\otimes k}){}\nonumber\\&&
     = (\bra{+}^{\otimes k-1})(\mathcal{F}(\rho_{\omega, k-1}(k)))(\ket{+}^{\otimes k-1})
    + 2\sqrt{\theta\overline{\theta}} (\bra{+}^{\otimes k-1}) (\mathcal{R}(\rho_{\omega, k-1}(k))) (\ket{+}^{\otimes k-1}).
\end{eqnarray}
\end{widetext}
We can write this as $M_{k}(\rho_{\omega, k}(k))$ where $M_{k}$ denotes the measurement operation $(\ket{+}^{\otimes k})$. This trivial notation will again help us appreciate the recursive nature of the evolution:

\begin{widetext}
\begin{eqnarray}\label{measurement_k}
    M_{k}(\rho_{\omega,  k}(k)) &= (\bra{+}^{\otimes k}) \rho_k(k) (\ket{+}^{\otimes k}),
    = \frac{1}{2}\left\{
    (f_{\rho} + 2\sqrt{\theta\overline{\theta}}r_{\rho})M_{k-1}(\rho_{\omega,  (k-1)}(k)) +  (f_{\mathbb{I}}+ 2\sqrt{\theta\overline{\theta}}r_{\mathbb{I}}) \right.\nonumber\\
    &\left.(\mathbb{I}_d \otimes\bra{+}^{\otimes k-1}\otimes\mathrm{Tr}_{d \times d}(\rho_{\omega,  k-1}(k))\otimes \ket{+}^{\otimes k-1}\right\}.
\end{eqnarray}
\end{widetext}

\noindent Using this generalization, we can show how to obtain the third iterations as an example recursively as follows:\\

\begin{eqnarray}
   &&M_3(\rho_{\omega, 3}(3)) = \frac{1}{2}(f_{\rho} + 2\sqrt{\theta\overline{\theta}})M_2(\rho_{\omega, 2}(3)){}\nonumber\\
   &&+ \frac{1}{2}(f_\mathbb{I} + 2\sqrt{\theta\overline{\theta}})\frac{\mathbb{I}_d}{d}(\bra{+}^{\otimes 2} \mathrm{Tr}_{d \times d}[\rho_{\omega, 2}(3)]\ket{+}^{\otimes 2}),{}\nonumber\\
   && M_2(\rho_{\omega, 2}(3)) = \frac{1}{2}(f_{\rho} + 2\sqrt{\theta\overline{\theta}})M_1(\rho_{\omega, 1}(3)){}\nonumber\\
   && + \frac{1}{2}(f_\mathbb{I} + 2\sqrt{\theta\overline{\theta}})\frac{\mathbb{I}_d}{d}(\bra{+} \mathrm{Tr}_{d \times d}[\rho_{\omega, 1}(3)]\ket{+}), {}\nonumber\\
   && M_1(\rho_{\omega, 1}(3)) = \frac{1}{2}(f_{\rho} + 2\sqrt{\theta\overline{\theta}})M_0(\rho_{\omega, 0}(3)){}\nonumber\\
   && + \frac{1}{2}(f_\mathbb{I} + 2\sqrt{\theta\overline{\theta}})\frac{\mathbb{I}_d}{d} \mathrm{Tr}_{d \times d}[\rho_{\omega, 0}(3)].
\end{eqnarray}
\noindent Here, $M_0(\rho_{\omega, 0}(3)) = \rho(3)$ and $\mathrm{Tr}_{d \times d}[\rho_{\omega, 0}(3)] = \mathrm{Tr}[\rho(3)] = 1$.

\subsubsection{Success Probability}
If we simplify and abstract out the expression for the unnormalized output state after measurement and postselection obtained by eqn. (\ref{measurement_k}) as: 
\begin{equation}
    \text{(coefficient of }\rho\text{)}P(k,0,d) + \frac{\text{(coefficient of }\frac{\mathbb{I}_d}{d}\text{)}}{d},  
\end{equation}
we can calculate the success probability in a similar way to the first framework, where success probability can be written as:

\begin{multline}
    \label{eqn:recipe_success_prob}
    \frac{\text{(coefficient of }\rho\text{)}}{\text{(coefficient of }\rho\text{)} + \text{(coefficient of }\frac{\mathbb{I}_d}{d}\text{)}}P(k,0,d)\\ 
    + \frac{\text{(coefficient of }\frac{\mathbb{I}_d}{d}\text{)}}{\text{(coefficient of }\rho\text{)} + \text{(coefficient of }\frac{\mathbb{I}_d}{d}\text{)}}\frac{1}{d},  
\end{multline}

\noindent Thus, we can write the success probability for the first three iterations in terms of the parameter $t$ as, 
\begin{widetext}
\begin{eqnarray}
&&P_{\omega}(1,(1-t),d) = P_\xi(1,(1-t),d) = 
    \left(f_\xi(t)\right)
    P(1,0,d)+\frac{1- f_\xi(t)}
    {d},
\end{eqnarray}
\begin{eqnarray}
&&P_{\omega}(2,(1-t),d)\nonumber\\
    &&=\frac{\left(\left(\frac{1-\sqrt{t}}{d}\right)^2 + 2t\right)^2
}{\left( \left(\left(\frac{1-\sqrt{t}}{d}\right)^2 + 2t\right)^2 + \left(1 + 2(1 - \sqrt{t})\sqrt{t} - t\right)\left(1 + \frac{1}{d^2}(1 - \sqrt{t})^2 + 2(1 - \sqrt{t})\sqrt{t} + 3t\right)\right)}P(2,0,d)\nonumber\\
&&\quad
    + \frac{\left(1 + 2(1 - \sqrt{t})\sqrt{t} - t\right)\left(1 + \frac{1}{d^2}(1 - \sqrt{t})^2 + 2(1 - \sqrt{t})\sqrt{t} + 3t\right)}{\left( \left(\left(\frac{1-\sqrt{t}}{d}\right)^2 + 2t\right)^2 + \left(1 + 2(1 - \sqrt{t})\sqrt{t} - t\right)\left(1 + \frac{1}{d^2}(1 - \sqrt{t})^2 + 2(1 - \sqrt{t})\sqrt{t} + 3t\right)\right)}\frac{1}{d}.\nonumber\\    
\end{eqnarray}
{\footnotesize
\begin{eqnarray}
    &&{\normalsize P_{\omega}(3,(1-t),d)}\nonumber\\
    &&=\frac{\left(\left(\frac{1-\sqrt{t}}{d}\right)^2 + 2t\right)^3
}{\left( \left(\left(\frac{1-\sqrt{t}}{d}\right)^2 + 2t\right)^3 + \left(1 + 2(1 - \sqrt{t})\sqrt{t} - t\right)\left(1 + \frac{1}{2d^2}(1 - \sqrt{t})^2 + 2(1 - \sqrt{t})\sqrt{t} + 3t + \left(\frac{1}{d^2}(1 - \sqrt{t})^2 + 2t\right)^2\right)
\right)}P(3,0,d)\nonumber\\
&&\quad
    + \frac{\left(1 + 2(1 - \sqrt{t})\sqrt{t} - t\right)\left(1 + \frac{1}{2d^2}(1 - \sqrt{t})^2 + 2(1 - \sqrt{t})\sqrt{t} + 3t + \left(\frac{1}{d^2}(1 - \sqrt{t})^2 + 2t\right)^2\right)
}{\left( \left(\left(\frac{1-\sqrt{t}}{d}\right)^2 + 2t\right)^3 + \left(1 + 2(1 - \sqrt{t})\sqrt{t} - t\right)\left(1 + \frac{1}{2d^2}(1 - \sqrt{t})^2 + 2(1 - \sqrt{t})\sqrt{t} + 3t + \left(\frac{1}{d^2}(1 - \sqrt{t})^2 + 2t\right)^2\right)
\right)}\frac{1}{d}.\nonumber\\
\end{eqnarray}}
\end{widetext}

\begin{figure*}[ht]
\noindent \begin{mdframed}
[
    linecolor=black,linewidth=0.5pt, frametitlerule=true,
    apptotikzsetting={\tikzset{mdfframetitlebackground/.append style={
        shade,left color=white, right color=gray!20}}}, 
    frametitlerulecolor=black,
    frametitlerulewidth=0.5pt, innertopmargin=\topskip,
    frametitle={Framework-2: Comparison of Success probability between the switch and non-switch scenario},
   ]
    \includegraphics[scale=0.9]{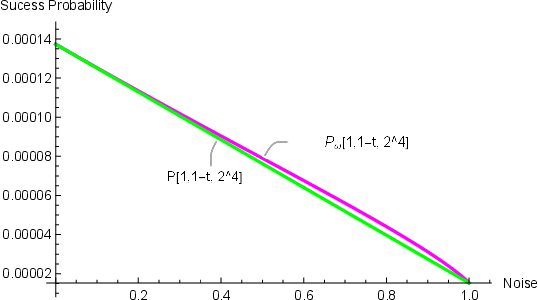}
    \includegraphics[scale=0.9]{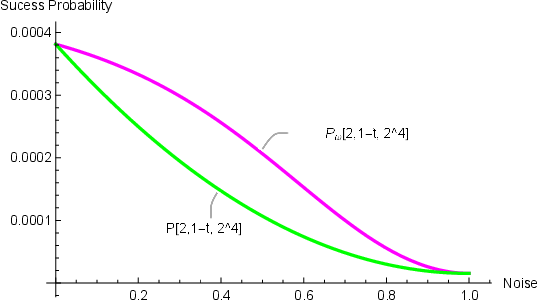}\\
    \includegraphics[scale=0.9]{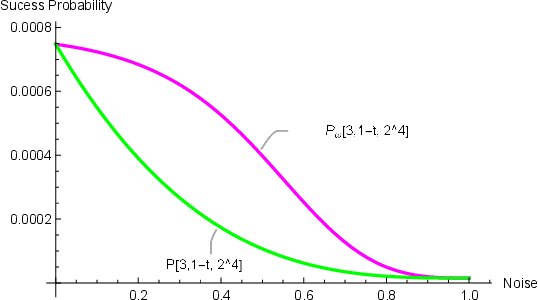}
        \begin{minipage}[b]{0.5\textwidth}
        \caption{ These plots show the effect of noise strength $(1-t)$ on the success probability of finding the target element in the search space in Noisy Grover's search algorithm. We are taking the search space to be $d = 2^4$, and thus the algorithm should stop at $k_Gr = \frac{\pi}{4}\sqrt{16} = \pi \approx 3$ iterations. The Plots from left to right show these variations for different iterations: k = 1 (top-left), k = 2 (top-right), and 3 (bottom-left). Here, the green curve represents the Success probability without using any switches, and the magenta curve represents the success probability on applying switches as described in fig.\ref{fig: noisy_grovers_framework2_last_iter}}
        \label{fig: F2_omega_Pvs(1-t)}
    \end{minipage}
    \end{mdframed}
\end{figure*}
\noindent Here, we plot the fig.(\ref{fig: F2_omega_Pvs(1-t)}) comparing the success probability of the switched channel framework 2 and the success probability of the noisy Grover iteration when there is no switch as the noise $(1-t)$ varies across $x$-axis. Just like fig.(\ref{fig: F1_xi}), it is again clear that the switch is giving an advantage in terms of restoring the success probability for the first and subsequent $k=2$ and $k=3$ (we take $d=2^4$ as an example). Ideally, the algorithm is supposed to stop after $k_{Gr}= \frac{\pi}{4}\sqrt{2^4} = \pi \approx 3$ iterations, and we can see how an increase in noise can drastically bring down the success probability from 1. This can be prevented using the switched channel framework 2 proposed in this section, which allows the system to tolerate more noise while keeping the success probability within an acceptable range.

\section{Conclusions}\label{sec:conclusions}

In conclusion, in this paper, we have explored the question if the use of indefinite causal order in quantum algorithms can be 
a promising approach for mitigating noise and improving performance.
We have shown that the indefinite causal order in noisy channels can significantly enhances the success probability of Grover's algorithm by mitigating the detrimental effects of noise. By allowing the quantum operations to exist in a superposition of different causal orders, this approach effectively distributes the noise across multiple pathways, reducing its overall impact on the system. This interference of various causal sequences helps average out noise, leading to higher success rate in finding the marked item. As a result, Grover's algorithm, which is highly sensitive to noise due to the delicate nature of quantum superposition, can operate with greater accuracy and efficiency. This novel technique opens up new possibilities for improving the performance of quantum algorithms, especially in near-term quantum devices where noise remains a significant challenge.

Our proposal may come in handy in a situation when there is no way of correcting fault-tolerant errors. In particular, our method becomes useful and handy  in case of computing with NISQ devices. We have presented two scenarios, one where the measurement takes place at every iteration and the other where the measurement is done at the final step.  In both methodologies, we can sustain the algorithm's efficacy in terms of success probability compared to the usual noisy scenario. In particular, the second methodology significantly outperforms the first one.

It is important to mention that we are not comparing our proposal to a
scenario where one can avoid the depolarising channel. Ideally, we can avoid noise, but our main objective  is to have a comparison between the scenarios where noise
acts directly and noise acts in the presence of indefinite causal order. It is known that if we can avoid the noise,
then,  we have nothing to worry about the noisy mitigation schemes in NISQ era. In future, it might be possible that in some physical scenario (like NISQ), noise may be present in quantum switches. In
such a scenario having a theoretical analysis like ours is valuable to enhance the success of quantum algorithms.

Since we have not used any specific property of Grover's search algorithm, the proposed scheme can be used for other iterative quantum algorithms. In future, one may be able to explore the effect of noise in the Variational Quantum Algorithms, 
quantum phase estimation algorithms \cite{chapeau-blondeau_indefinite_2023} and others alike. By leveraging the notion of indefinite causal order, this may provide new pathways for designing quantum algorithms that are more resilient to noise and other imperfections in real-world quantum systems. We hope that our results will find useful applications in near-term quantum devices where noise remains a major bottleneck for scaling quantum computing technologies.

\noindent \textit{Acknowledgement:} SS acknowledges useful discussions with Dr. Shantanav Chakraborty and Dr. Siddhartha Das 

\bibliography{references}
\onecolumngrid

\section{Appendix}\label{sec:appendix}   
\subsection{Calculations for the first iteration (k = 1) in noisy Grover's Search with Algorithm with quantum  switch}\label{appendix:q_switch_k=1_calc}
\noindent For $k = 1$, we have $\mathcal{S} (\mathcal{D}_{\sqrt{t}} , \mathcal{D}_{\sqrt{t}}) (\rho \otimes \rho_c)$. Now, by expanding the inputs in the  expression, we have\\
1. The depolarising channel noise represented as $\mathcal{D}_{\sqrt{t}}$
\begin{eqnarray}\label{eqn: kraus_unitary_operators_calc}
    \mathcal{D}_{\sqrt{t}}(\rho) &&= \sqrt{t} \rho + (1 - \sqrt{t}) \mathrm{Tr}[\rho] \dfrac{\mathbb{I}_d}{d}\nonumber\\&&
    = \sqrt{t}\rho + (1 - \sqrt{t})\dfrac{1}{d^2}\mathlarger{\mathlarger{\sum\limits}}_{i = 1}^{d^2} U_i \rho U_i^\dagger\qquad\text{using, } \dfrac{1}{d^2}\sum_{i = 1}^{d^2}U_i \rho U_{i}^{\dagger} = \mathrm{Tr}[\rho]\dfrac{\mathbb{I}_d}{d}\nonumber\\&&
    = \dfrac{(1-\sqrt{t})}{d^2}\Bigg (\dfrac{\sqrt{t}d^2}{1-\sqrt{t}} \mathbb{I}_d \rho \mathbb{I}_d^\dagger + \mathlarger{\mathlarger{\sum\limits}}_{i = 1}^{d^2} U_i \rho U_i^\dagger \Bigg){}\nonumber\\&&
    = \dfrac{(1-\sqrt{t})}{d^2}\Bigg ( U_0 \rho U_0^\dagger + \mathlarger{\mathlarger{\sum\limits}}_{i = 1}^{d^2}   U_i   \rho   U_i^\dagger \Bigg )\qquad\text{taking, } U_0 = \sqrt{\dfrac{\sqrt{t}d^2}{1-\sqrt{t}}}   \mathbb{I}_d\nonumber\\&&
    = \dfrac{(1-\sqrt{t})}{d^2} \mathlarger{\mathlarger{\sum\limits}}_{i = 0}^{d^2}   U_i   \rho   U_i^\dagger
\end{eqnarray}
Using, $K_0 = \sqrt{\dfrac{(1-\sqrt{t})}{d^2}}U_0 = \sqrt[4]{t}\mathbb{I}_d$ and $K_i = \sqrt{\dfrac{(1-\sqrt{t})}{d^2}}U_i$ \,where \{$i = 1,2,\ldots d^2$\}. We have the set of Kraus operators $\{K_i\}$ for $\mathcal{D}_{\sqrt{t}}$ 
\begin{equation} \label{def:dep_kraus_sqrt(t)}
  \mathcal{D}_{\sqrt{t}}(\rho)= \mathlarger{\mathlarger{\sum\limits}}_{i = 0}^{d^2} K_i \rho K_i^\dagger  
\end{equation}\\
2. The control qubit represented as denoted by $\rho_c$ \begin{eqnarray}
    \rho_c &&= \ket{c}\bra{c}\ \qquad \text{where, }\ket{c}=\sqrt{\theta}\ket{0} + \sqrt{\overline{\theta}}\ket{1}{}\nonumber\\&&
=\theta\ket{0}\bra{0}+\sqrt{\theta\overline{\theta}}\ket{0}\bra{1}+\sqrt{\theta\overline{\theta}}\ket{1}\bra{0}+\overline{\theta}\ket{1}\bra{1}
\end{eqnarray}
3. The input search space after the application of the first Grover iteration, consisting of $n$-qubits $\equiv$ $2^n$ dimensional qudit system denoted by $\rho(1)$.\\

\noindent We also note the following for later use 
\begin{equation}\label{prop:TrUVU=V}
\frac{1}{d} \sum_{i=1}^{d^2} \mathrm{Tr}[U_i^\dagger V] U_i = V,
\end{equation}

\begin{equation}\label{prop:UVU=TrV}
\frac{1}{d^2} \sum_{i=1}^{d^2} U_i V U_i^\dagger = \mathrm{Tr}[V] \frac{\mathbb{I}_d}{d},
\end{equation}
Since the \( d^2 \) unitary operators \( U_i \) are orthogonal to each other, they establish an orthonormal basis with respect to the Hilbert-Schmidt inner product. This basis is applicable to any \( d \)-dimensional linear operator \( V \).\\
We can express the Kraus operators of the switch $\mathcal{S}$ \ref{def:switch_kraus_0}:
\begin{eqnarray}
 \mathcal{W}_{ij} &&= K_i K_j \otimes \ket{0}\bra{0} + K_j K_i \otimes \ket{1}\bra{1}{}\nonumber\\&& = \dfrac{(1-\sqrt{t})}{d^2}\Bigg (U_i U_j \otimes \ket{0}\bra{0} + U_j U_i \otimes \ket{1}\bra{1}\Bigg )  \text{for } i,j = 0,1,2\ldots d^2
\end{eqnarray}
\begin{eqnarray}
  \mathcal{S} (\mathcal{D}_{\sqrt{t}}, \mathcal{D}_{\sqrt{t}})(\rho(1) \otimes \rho_{c}) 
        &&= \mathlarger{\mathlarger{\sum\limits}}_{i, \,j = 0}^{d^2} \mathcal{W}_{ij}(\rho(1) \otimes \rho_{c}) \mathcal{W}_{ij}^{\dagger}\nonumber\\
        &&= \mathlarger{\mathlarger{\sum\limits}}_{i, j = 0}^{d^2} \Bigl\{ \bigl(K_i K_j\otimes \ket{0} \bra{0} + K_j K_i \otimes \ket{1} \bra{1}\bigl)\bigl(\rho(1) \otimes \rho_{c}\bigl)\nonumber\\
        &&\qquad\bigl(K_i K_j\otimes \ket{0} \bra{0} + K_j K_i \otimes \ket{1} \bra{1}\bigl)^{\dagger}\Bigl\}{}\\
        &&= \mathlarger{\mathlarger{\sum\limits}}_{i, j = 0}^{d^2}\Bigl\{ \bigl (K_i K_j  \rho(1)  K_j^{\dagger} K_i^{\dagger} \otimes \ket{0}\bra{0}  \rho_{c}  \ket{0}\bra{0}\bigl )  {}\nonumber\\ 
        &&\qquad+  \bigl (K_i K_j  \rho(1)  K_j^{\dagger} K_i^{\dagger} \otimes \ket{0}\bra{0}  \rho_{c_1}  \ket{1}\bra{1}\bigl ){}\nonumber\\
        &&\qquad+ \bigl (K_i K_j  \rho(1)  K_j^{\dagger} K_i^{\dagger} \otimes \ket{1}\bra{1}  \rho_{c_1}  \ket{0}\bra{0}\bigl ){}\nonumber\\
        &&\qquad+  \bigl (K_i K_j  \rho(1)  K_j^{\dagger} K_i^{\dagger} \otimes \ket{1}\bra{1}  \rho_{c_1}  \ket{1}\bra{1}\bigl ) \Bigl\}\\ 
  \mathcal{S} (\mathcal{D}_{\sqrt{t}}, \mathcal{D}_{\sqrt{t}})(\rho(1) \otimes \rho_{c_1}) 
        &&=  \mathlarger{\mathlarger{\sum\limits}}_{i, j = 0}^{d^2}\biggl \{ \bigl (K_i K_j  \rho(1)  K_j^{\dagger} K_i^{\dagger} \otimes \theta \ket{0}\bra{0}\bigl )\nonumber\\  
        &&\qquad+  \bigl (K_i K_j  \rho(1)  K_j^{\dagger} K_i^{\dagger} \otimes \sqrt{\theta\overline{\theta}} \ket{0}\bra{1}\bigl ) {}\nonumber\\
        &&\qquad+ \bigl (K_i K_j  \rho(1)  K_j^{\dagger} K_i^{\dagger} \otimes \sqrt{\theta\overline{\theta}} \ket{1}\bra{0}\bigl )\nonumber\\ 
        &&\qquad+ \bigl (K_i K_j  \rho(1)  K_j^{\dagger} K_i^{\dagger} \otimes \overline{\theta} \ket{1}\bra{1}\bigl ) \biggl \}
\end{eqnarray}
\ref{def:dep_kraus_sqrt(t)} gives the Kraus operators. We will break this equation into four parts: \\
1. After fixing both $i = j = 0$ , we have
\begin{eqnarray}
    &&\mathlarger{\mathlarger{\sum\limits}}_{i = 1}^{d^2}\mathlarger{\mathlarger{\sum\limits}}_{j = 1}^{d^2} \Biggl \{ \Biggl (\sqrt{\dfrac{1-\sqrt{t}}{d^2}}U_i \sqrt{\dfrac{1-\sqrt{t}}{d^2}}U_j \rho(1) \sqrt{\dfrac{1-\sqrt{t}}{d^2}}U_j^{\dagger}\sqrt{\dfrac{1-\sqrt{t}}{d^2}}U_i^{\dagger} \otimes  \theta \ket{0}\bra{0}\Biggl ) {}\nonumber\\
    &&\qquad+ \Biggl (\sqrt{\dfrac{1-\sqrt{t}}{d^2}}U_i \sqrt{\dfrac{1-\sqrt{t}}{d^2}}U_j \rho(1) \sqrt{\dfrac{1-\sqrt{t}}{d^2}}U_j^{\dagger}\sqrt{\dfrac{1-\sqrt{t}}{d^2}}U_i^{\dagger} \otimes  \sqrt{\theta\overline{\theta}} \ket{0}\bra{1}\Biggl ) {}\nonumber\\
    &&\qquad+ \Biggl (\sqrt{\dfrac{1-\sqrt{t}}{d^2}}U_i \sqrt{\dfrac{1-\sqrt{t}}{d^2}}U_j \rho(1) \sqrt{\dfrac{1-\sqrt{t}}{d^2}}U_j^{\dagger}\sqrt{\dfrac{1-\sqrt{t}}{d^2}}U_i^{\dagger} \otimes  \sqrt{\theta\overline{\theta}} \ket{1}\bra{0}\Biggl ) {}\nonumber\\
    &&\qquad+ \Biggl (\sqrt{\dfrac{1-\sqrt{t}}{d^2}}U_i \sqrt{\dfrac{1-\sqrt{t}}{d^2}}U_j \rho(1) \sqrt{\dfrac{1-\sqrt{t}}{d^2}}U_j^{\dagger}\sqrt{\dfrac{1-\sqrt{t}}{d^2}}U_i^{\dagger} \otimes  \overline{\theta} \ket{1}\bra{1}\Biggl )
    \Biggl \}{}\nonumber\\
\end{eqnarray}
\begin{eqnarray}
        &&= (1-\sqrt{t})^2\Biggl \{\dfrac{1}{d^2 }\mathlarger{\mathlarger{\sum\limits}}_{i = 1}^{d^2}U_i\Biggl (\dfrac{1}{d^2}\mathlarger{\mathlarger{\sum\limits}}_{j = 1}^{d^2} U_j   \rho(1)  U_j^{\dagger}\Biggl ) U_i^{\dagger} \otimes \theta \ket{0}\bra{0}\nonumber\\ 
        &&\qquad+\dfrac{1}{d^2 }\mathlarger{\mathlarger{\sum\limits}}_{j = 1}^{d^2}\Biggl (\dfrac{1}{d^2}\mathlarger{\mathlarger{\sum\limits}}_{i = 1}^{d^2} U_i   \biggl(U_j \rho(1)\biggl)  U_i^{\dagger}\Biggl ) U_j^{\dagger} \otimes \sqrt{\theta\overline{\theta}} \ket{0}\bra{1}{}\nonumber\\
        &&\quad+\dfrac{1}{d^2 }\mathlarger{\mathlarger{\sum\limits}}_{i = 1}^{d^2}\Biggl (\dfrac{1}{d^2}\mathlarger{\mathlarger{\sum\limits}}_{j = 1}^{d^2} U_j   \biggl(U_i \rho(1)\biggl)  U_j^{\dagger}\Biggl ) U_i^{\dagger} \otimes \sqrt{\theta\overline{\theta}} \ket{1}\bra{0}\nonumber\\ 
        &&\qquad+\dfrac{1}{d^2 }\mathlarger{\mathlarger{\sum\limits}}_{j = 1}^{d^2}U_j\Biggl (\dfrac{1}{d^2}\mathlarger{\mathlarger{\sum\limits}}_{i = 1}^{d^2} U_i   \rho(1)  U_i^{\dagger}\Biggl ) U_j^{\dagger} \otimes \overline{\theta} \ket{1}\bra{1}\Biggl \}{}
\end{eqnarray}
\begin{eqnarray}
    &&= (1-\sqrt{t})^2\Biggl \{\dfrac{1}{d^2 }\mathlarger{\mathlarger{\sum\limits}}_{i = 1}^{d^2}U_i \Biggl (\mathrm{Tr}\bigl(\rho(1)\bigl) \dfrac{\mathbb{I}_d}{d}\Biggl ) U_i^{\dagger} \otimes \theta\ket{0}\bra{0} + \dfrac{1}{d^2}\mathlarger{\mathlarger{\sum\limits}}_{j = 1}^{d^2} \Biggl (\mathrm{Tr}\bigl(U_j\rho(1)\bigl)\dfrac{\mathbb{I}_d}{d}U_j^{\dagger}\Biggl ) \otimes \sqrt{\theta\overline{\theta}}\ket{0}\bra{1}{}\nonumber\\
        &&+\quad\dfrac{1}{d^2}\mathlarger{\mathlarger{\sum\limits}}_{i = 1}^{d^2} \Biggl (\mathrm{Tr}\bigl(U_i\rho(1)\bigl)\dfrac{\mathbb{I}_d}{d}U_i^{\dagger}\Biggl ) \otimes  \sqrt{\theta\overline{\theta}}\ket{1}\bra{0}+\dfrac{1}{d^2}\mathlarger{\mathlarger{\sum\limits}}_{j = 1}^{d^2}U_j \Biggl (\mathrm{Tr}\bigl(\rho(1)\bigl)\dfrac{\mathbb{I}_d}{d}\Biggl ) U_j^{\dagger} \otimes \overline{\theta}\ket{1}\bra{1}
        \Biggl\}{}
\end{eqnarray}
\begin{eqnarray}
    &&= (1-\sqrt{t})^2\Biggl\{\mathrm{Tr}[\rho]
        \dfrac{\mathbb{I}_d}{d}\otimes\theta\ket{0}\bra{0}
        +\dfrac{1}{d^2}\rho(1)\otimes \sqrt{\theta\overline{\theta}} \ket{0}\bra{1}{}\nonumber\\
        &&+\dfrac{1}{d^2}\rho(1)\otimes \sqrt{\theta\overline{\theta}} \ket{1}\bra{0}
        + \mathrm{Tr}[\rho]
        \dfrac{\mathbb{I}_d}{d}\otimes\overline{\theta}\ket{1}\bra{1} \Biggl\}
        \label{eqn:switch_i=0,j=0}
\end{eqnarray}
2. fix $i = 0, j \neq 0$.
    \begin{eqnarray}
        &&= \mathlarger{\mathlarger{\sum\limits}}_{j = 1}^{d^2} \Biggl\{\sqrt[4]{t} \mathbb{I}_d\sqrt{\dfrac{1-\sqrt{t}}{d^2}}U_j \rho(1) \sqrt{\dfrac{1-\sqrt{t}}{d^2}}U_j^{\dagger}\sqrt[4]{t} \mathbb{I}_d \otimes  \theta \ket{0}\bra{0}{}\nonumber\\
        &&\quad+ \sqrt[4]{t} \mathbb{I}_d\sqrt{\dfrac{1-\sqrt{t}}{d^2}}U_j \rho(1)\sqrt[4]{t} \mathbb{I}_d \sqrt{\dfrac{1-\sqrt{t}}{d^2}}U_j^{\dagger} \otimes  \sqrt{\theta\overline{\theta}} \ket{0}\bra{1}{}\nonumber\\
        &&\quad+ \sqrt{\dfrac{1-\sqrt{t}}{d^2}}U_j \sqrt[4]{t} \mathbb{I}_d \rho(1)\sqrt{\dfrac{1-\sqrt{t}}{d^2}}U_j^{\dagger} \sqrt[4]{t}\otimes  \sqrt{\theta\overline{\theta}} \ket{1}\bra{0}{}\nonumber\\
        &&\quad+ \sqrt{\dfrac{1-\sqrt{t}}{d^2}}U_j \sqrt[4]{t} \mathbb{I}_d \rho(1) \sqrt[4]{t} \mathbb{I}_d\sqrt{\dfrac{1-\sqrt{t}}{d^2}} \otimes  \overline{\theta}\ket{1}\bra{0}\Biggl \}{}\\
        &&=\mathlarger{\mathlarger{\sum\limits}}_{j=1}^{d^2}\Biggl(\sqrt{\dfrac{\sqrt{t}(1-\sqrt{t})}{d^2}}\Biggl)^{2}\Biggl(U_j \rho(1) U_j^{\dagger} \otimes \theta\ket{0}\bra{0} + U_j \rho(1) U_j^{\dagger} \otimes \overline{\theta}\ket{1}\bra{1}{}\nonumber\\
        &&= \Biggl(\dfrac{\sqrt{t}(1-\sqrt{t})}{d^2}\Biggl)\mathlarger{\mathlarger{\mathlarger{\sum\limits}}}_{j=1}^{d^2}\Biggl(U_j \rho(1) U_j^{\dagger}\nonumber\\ &&\qquad\otimes \biggl(\theta\ket{0}\bra{0} + \overline{\theta}\ket{1}\bra{1} + \sqrt{\theta\overline{\theta}}\biggl(\ket{0}\bra{1} + \ket{1}\bra{0} \biggl) \biggl) \Biggl){}\\
        &&= \sqrt{t}\bigl(1-\sqrt{t}\bigl) \dfrac{1}{d^2} \mathlarger{\mathlarger{\sum\limits}}_{j = 1}^{d^2}\Biggl(U_J \rho(1) U_J^{\dagger} \otimes \rho_c\Biggl){} = \sqrt{t}\bigl(1-\sqrt{t}\bigl) \mathrm{Tr}[\rho] \dfrac{\mathbb{I}_d}{d} \otimes \rho_c \label{eqn:switch_i_n=_0,j=0}
        \end{eqnarray}
3. This is symmetric to \ref{eqn:switch_i_n=_0,j=0} and will evaluate to the same value
\begin{equation} 
    = \sqrt{t}\bigl(1-\sqrt{t}\bigl) \mathrm{Tr}[\rho] \dfrac{\mathbb{I}_d}{d} \otimes \rho_c \label{eqn:switch_i=0,j_n=_0}
\end{equation}
4. Finally, fix $i \neq 0, j \neq 0$
\begin{eqnarray} 
    = \sqrt[4]{t} \mathbb{I}_d\sqrt[4]{t} \mathbb{I}_d\bigl( \rho  \otimes \rho_c\bigl)\sqrt[4]{t} \mathbb{I}_d\sqrt[4]{t} \mathbb{I}_d\nonumber\\
    = t \bigl(\rho \otimes \rho_C\bigl)\label{eqn:switch_i_n=_0,j_n=_0}
\end{eqnarray}
Hence, \ref{eqn:switch_i=0,j=0}+\ref{eqn:switch_i_n=_0,j=0}+\ref{eqn:switch_i=0,j_n=_0}+\ref{eqn:switch_i_n=_0,j_n=_0}
\begin{equation}\label{eqn:switch_k=1_iter}
    \begin{aligned}
        &\mathcal{S} (\mathcal{D}_{\sqrt{t}}, \mathcal{D}_{\sqrt{t}})(\rho(1) \otimes \rho_{c_1}) =\\ 
        &(1-\sqrt{t})^2\Biggl (\mathrm{Tr}[\rho]\dfrac{\mathbb{I}_d}{d}   \otimes  \biggl(\theta\ket{0}\bra{0} + \overline{\theta}\ket{1}\bra{1}\biggl) +  \dfrac{\rho}{d^2}  \otimes \biggl(\sqrt{\theta\overline{\theta}}\biggl(\ket{0}\bra{1} + \ket{1}\bra{0}\biggl)\biggl) \Biggl)\\  
        &\qquad+ 2\sqrt{t}\bigl(1-\sqrt{t}\bigl) \mathrm{Tr}[\rho] \biggl(\dfrac{\mathbb{I}_d}{d} \otimes \rho_c\biggl) + t \bigl(\rho \otimes \rho_C\bigl)
    \end{aligned}
\end{equation}
Collecting Terms, this output can also be written as:
\begin{eqnarray*}
&&\begin{aligned}
    =& \Biggl[\Biggl [ \biggl(1- \sqrt{t}\biggl)^2\Biggl( \mathrm{Tr}[\rho]\dfrac{\mathbb{I}_d}{d} \otimes \biggl(\theta\ket{0}\bra{0} + \overline{\theta}\ket{1}\bra{1}\biggl) + \dfrac{\rho}{d^2} \otimes \sqrt{\theta\overline{\theta}}\biggl(\ket{0}\bra{1} + \ket{1}\bra{0}\biggl)\Biggl)\Biggl]\\
    &\qquad+ \Biggl [2\sqrt{t}\biggl(1-\sqrt{t}\biggl) \mathrm{Tr}[\rho]\dfrac{\mathbb{I}_d}{d} \otimes \biggl(\theta\ket{0}\bra{0} + \sqrt{\theta\overline{\theta}}\biggl(\ket{0}\bra{1}+\ket{1}\bra{0}\biggl) + \overline{\theta}\ket{1}\bra{1}\Biggl]\\
    &\qquad+ \Biggl[t\rho \otimes \biggl(\theta\ket{0}\bra{0} + \sqrt{\theta\overline{\theta}}\biggl(\ket{0}\bra{1}+\ket{1}\bra{0}\biggl) + \overline{\theta}\ket{1}\bra{1}\Biggl]\Biggl]
\end{aligned}\\
&&\begin{aligned}
    =& \biggl(1-\sqrt{t}\biggl)^2\Biggl(\mathrm{Tr}[\rho]\dfrac{\mathbb{I}_d}{d} \otimes \biggl(\theta\ket{0}\bra{0} + \overline{\theta}\ket{1}\bra{1}\biggl)\Biggl) + 2\sqrt{t}\biggl(1-\sqrt{t}\biggl)\Biggl(\mathrm{Tr}[\rho]\dfrac{\mathbb{I}_d}{d} \otimes \biggl(\theta\ket{0}\bra{0} + \overline{\theta}\ket{1}\bra{1}\biggl)\Biggl)\\
    &\qquad+ t\Biggl(\rho \otimes \biggl(\theta\ket{0}\bra{0} + \overline{\theta}\ket{1}\bra{1}\biggl)\Biggl) + \dfrac{(1-\sqrt{t})^2}{d^2}\Biggl(\rho \otimes \sqrt{\theta\overline{\theta}}\biggl(\ket{0}\bra{1}+\ket{1}\bra{0}\biggl)\Biggl) \\
    &\qquad+ t\Biggl(\rho \otimes \sqrt{\theta\overline{\theta}}\biggl(\ket{0}\bra{1}+\ket{1}\bra{0}\biggl)\Biggl)
    + 2\sqrt{t}\biggl(1-\sqrt{t}\biggl)\Biggl(\mathrm{Tr}[\rho]\dfrac{\mathbb{I}_d}{d} \otimes \sqrt{\theta\overline{\theta}}\biggl(\ket{0}\bra{1}+\ket{1}\bra{0}\biggl)\Biggl)
\end{aligned}\\
&&\begin{aligned}
    =& \Biggl((1-\sqrt{t})^2+2\sqrt{t}(1-\sqrt{t})\Biggl) \mathrm{Tr}[\rho]\dfrac{\mathbb{I}_d}{d} \otimes \Biggl(\theta\ket{0}\bra{0} + \overline{\theta}\ket{1}\bra{1}\Biggl)  +  (t)\rho \otimes  \Biggl(\theta\ket{0}\bra{0} + \overline{\theta}\ket{1}\bra{1}\Biggl)\\
    &\qquad+ \Biggl(\biggl(\dfrac{1-\sqrt{t}}{d}\biggl)^2+ t\Biggl)\Biggl(\rho \otimes \sqrt{\theta\overline{\theta}}\biggl(\ket{0}\bra{1}+\ket{1}\bra{0}\biggl) \Biggl)\\ 
    &\qquad+ 2\sqrt{t}\biggl(1-\sqrt{t}\biggl)\Biggl(\mathrm{Tr}[\rho]\dfrac{\mathbb{I}_d}{d} \otimes  \sqrt{\theta\overline{\theta}}\biggl(\ket{0}\bra{1}+\ket{1}\bra{0}\biggl)\Biggl)
\end{aligned}\\
  &&\begin{aligned}
    &\mathcal{S} (\mathcal{D}_{\sqrt{t}}, \mathcal{D}_{\sqrt{t}})(\rho(1) \otimes \rho_{c_1}) =
    \Biggl(\bigl(1-t\bigl)\mathrm{Tr}[\rho]\dfrac{\mathbb{I}_d}{d}+t\rho\Biggl)\otimes\Biggl(\theta\ket{0}\bra{0}+\overline{\theta}\ket{1}\bra{1}\Biggl)\\
    &+\left\{\Biggl(\biggl(\dfrac{(1-\sqrt{t})}{d}\biggl)^2+t\Biggl)\rho + 2\sqrt{t}\bigl(1-\sqrt{t}\bigl)\Biggl(\mathrm{Tr}[\rho]\dfrac{\mathbb{I}_d}{d}\Biggl)\right\}\otimes\sqrt{\theta\overline{\theta}}\biggl(\ket{0}\bra{1}+\ket{1}\bra{0} \biggl)
\end{aligned}  
\end{eqnarray*}
Using notations introduced from eqn.(\ref{def: f(t)r(t)_notation}) to (\ref{def: F(rho)R(rho)_notation})

\begin{eqnarray}
    &&\mathcal{S} (\mathcal{D}_{\sqrt{t}}, \mathcal{D}_{\sqrt{t}})(\rho(1) \otimes \rho_{c_1})\nonumber\\
    &&\qquad= \Biggl(f_{\mathbb{I}}\mathrm{Tr}[\rho]\dfrac{\mathbb{I}_d}{d}+t\rho\Biggl)\otimes\Biggl(\theta\ket{0}\bra{0}+\overline{\theta}\ket{1}\bra{1}\Biggl)\nonumber\\
    &&\quad\qquad+\left\{r_{\rho}\rho + r_{\mathbb{I}}\mathrm{Tr}[\rho]\dfrac{\mathbb{I}_d}{d}\right\}\otimes\sqrt{\theta\overline{\theta}}\biggl(\ket{0}\bra{1}+\ket{1}\bra{0} \biggl)\nonumber\\
    &&\qquad= \begin{bmatrix}
    \theta\Biggl(f_{\rho}\rho + f_{\mathbb{I}} \mathrm{Tr}[\rho]\dfrac{\mathbb{I}_d}{d}\Biggl)
        & \sqrt{\theta\overline{\theta}}\Biggl(r_{\rho} \rho + r_{\mathbb{I}} \mathrm{Tr}[\rho]\dfrac{\mathbb{I}_d}{d}\Biggl)\\
    \sqrt{\theta\overline{\theta}}\Biggl(r_{\rho} \rho + r_{\mathbb{I}} \mathrm{Tr}[\rho]\dfrac{\mathbb{I}_d}{d}\Biggl) 
        & \overline{\theta}\Biggl(f_{\rho}\rho + f_{\mathbb{I}} \mathrm{Tr}[\rho]\dfrac{\mathbb{I}_d}{d}\Biggl)
    \end{bmatrix}\\    
    &&\qquad= \begin{bmatrix}
        \theta \mathcal{F}(\rho(1))
        & \sqrt{\theta\overline{\theta}} \mathcal{R}(\rho(1))\\
        \sqrt{\theta\overline{\theta}}\mathcal{R}(\rho(1)) 
        & \overline{\theta} \mathcal{F}(\rho(1)) 
        \end{bmatrix} \label{eqn:switch_k=1_iter_blockmatrix}       
\end{eqnarray}

Thus we have 3 ways \ref{eqn:switch_k=1_iter}, \ref{eqn:switch_k=1_iter_rearranged}, \ref{eqn:switch_k=1_iter_blockmatrix} to write the output for $\mathcal{S} (\mathcal{D}_{\sqrt{t}}, \mathcal{D}_{\sqrt{t}})(\rho(1) \otimes \rho_{c_1})$ 

\subsection{Measurement of the system} 
\begin{equation}
    \rho_{\xi}(1)=\frac{N}{M}=\frac{\bra{\pm}\mathcal{S}(\mathcal{D}_{\sqrt{t}}, \mathcal{D}_{\sqrt{t}})(\rho(1)\otimes\rho_{c})\ket{\pm}}{\mathrm{Tr}\left[\bra{\pm}\mathcal{S}(\mathcal{D}_{\sqrt{t}}, \mathcal{D}_{\sqrt{t}})(\rho(1)\otimes\rho_{c})\ket{\pm}\right]}
\end{equation}
Calculating the numerator, N:
\begin{eqnarray}
    &&\bra{\pm}\Biggl[\left\{t\rho(1) + (1-t)\mathrm{Tr}[\rho(1)]\frac{\mathbb{I}_d}{d}\right\} \otimes\left(\theta\ket{0}\bra{0} + \overline{\theta}\ket{1}\bra{1}\right)\nonumber\\
    &&\quad+\left\{\left(\left(\frac{1-\sqrt{t}}{d}\right)^2+t\right)\rho(1) + 2\sqrt{t}(1-\sqrt{t})\mathrm{Tr}[\rho(1)]\frac{\mathbb{I}_d}{d}\right\} \otimes\sqrt{\theta\overline{\theta}}\left(\ket{0}\bra{1}+\ket{1}\bra{0} \right)\Biggr]\ket{\pm}\\
    &&=\frac{1}{2}\left(t\rho(1) + (1-t)\mathrm{Tr}[\rho(1)]\frac{\mathbb{I}_d}{d}\right) + \frac{1}{2} \left(\pm2\sqrt{\theta\overline{\theta}}\left\{\left(\left(\frac{1-\sqrt{t}}{d}\right)^2+t\right)\rho(1) + 2\sqrt{t}(1-\sqrt{t})\mathrm{Tr}[\rho(1)]\frac{\mathbb{I}_d}{d}\right\}\right)\nonumber\\
    &&=\frac{1}{2}\left\{t\rho(1) + (1-t)\mathrm{Tr}[\rho(1)]\frac{\mathbb{I}_d}{d} \pm2\sqrt{\theta\overline{\theta}}\left(\left(\frac{1-\sqrt{t}}{d}\right)^2+t\right)\rho(1) \pm2\sqrt{\theta\overline{\theta}} \left(2\sqrt{t}(1-\sqrt{t})\mathrm{Tr}[\rho(1)]\frac{\mathbb{I}_d}{d}\right)\right\}\nonumber\\
    &&=\frac{1}{2}\left\{\left(t \pm2\sqrt{\theta\overline{\theta}}\left(\left(\frac{1-\sqrt{t}}{d}\right)^2+t\right)\right)\rho(1) +  \left((1-t) \pm2\sqrt{\theta\overline{\theta}} 2\sqrt{t}(1-\sqrt{t})\right)\mathrm{Tr}[\rho(1)]\frac{\mathbb{I}_d}{d}\right\}    
\end{eqnarray}
Calculating the denominator, M:
\begin{eqnarray}
    && \mathrm{Tr}\left[\frac{1}{2}\left(\left(\pm2\sqrt{\theta\overline{\theta}}(\frac{(1-\sqrt{t})^{2}}{d^{2}} + t) + t\right)\rho(1)
    + \left((1-t)\pm2\sqrt{\theta\overline{\theta}}(2\sqrt{t}(1-\sqrt{t}))\right)\mathrm{Tr}[\rho(1)]\frac{\mathbb{I}_d}{d}\right)\right]\\
    &&= \frac{1}{2}\left\{\mathrm{Tr}\left[\left(\pm2\sqrt{\theta\overline{\theta}}(\frac{(1-\sqrt{t})^{2}}{d^{2}} + t) + t\right)\rho(1)\right]+ \mathrm{Tr}\left[\left((1-t)\pm2\sqrt{\theta\overline{\theta}}(2\sqrt{t}(1-\sqrt{t}))\right)\mathrm{Tr}[\rho(1)]\frac{\mathbb{I}_d}{d}\right]\right\}\nonumber\\
    &&= \frac{1}{2}\left\{\left(\pm 2\sqrt{\theta\overline{\theta}}(\frac{(1-\sqrt{t})^{2}}{d^{2}} + t) + t\right)+ \left((1-t)\pm 2\sqrt{\theta\overline{\theta}}(2\sqrt{t}(1-\sqrt{t}))\right)\right\}\qquad
\end{eqnarray}
Because $\mathrm{Tr}[\rho(1)] = \mathrm{Tr}[\frac{\mathbb{I}_d}{d}] = 1$. Thus, the density matrix after measurement in Fourier basis is:
\begin{eqnarray}
\frac{
\frac{1}{2}\left\{\left(t \pm2\sqrt{\theta\overline{\theta}}\left(\left(\frac{1-\sqrt{t}}{d}\right)^2+t\right)\right)\rho(1) +  \left((1-t) \pm2\sqrt{\theta\overline{\theta}} 2\sqrt{t}(1-\sqrt{t})\right)\mathrm{Tr}[\rho(1)]\frac{\mathbb{I}_d}{d}\right\}}{
\frac{1}{2}\left\{\left(\pm2\sqrt{\theta\overline{\theta}}\frac{(1-\sqrt{t})^{2}}{d^{2}} + t) + t\right)+ \left((1-t)\pm 2\sqrt{\theta\overline{\theta}}(2\sqrt{t}(1-\sqrt{t}))\right)\right\}}   
\end{eqnarray}
For $\theta = 0$ or $(1-\theta) = \overline{\theta} = 0$

\begin{eqnarray*}
    =& \frac{1}{2}\left(t\rho(1)+(1-t)\frac{\mathbb{I}_d}{d})\right)
\end{eqnarray*}
For maximum superposition in the switch, $\theta = (1-\theta) = \overline{\theta} = \frac{1}{2}$. Taking the $|+\rangle$ component of the measurement.  \textbf{Post Selection} and Correcting based on the + or - part of the measurement
\begin{eqnarray}
&&\frac{\left(2t + \left(\frac{1-\sqrt{t}}{d}\right)^2\right)\rho(1) +  \left((1-t) + 2\sqrt{t}(1-\sqrt{t})\right)\mathrm{Tr}[\rho(1)]\frac{\mathbb{I}_d}{d}}{\left(\frac{(1-\sqrt{t})^{2}}{d^{2}} + 2t\right)+ \left((1-t)+(2\sqrt{t}(1-\sqrt{t}))\right)}\\
&&= f_\xi(t)\rho(1) + (1-f_\xi(t)) \mathrm{Tr}[\rho(1)]\frac{\mathbb{I}_d}{d}
\end{eqnarray}
\begin{equation}\text{where, }
    f_\xi(t)= \frac{\left(\frac{1-\sqrt{t}}{d}\right)^2 + 2t}{\left(\frac{1+(t-2\sqrt{t})(1-d^2)}{d^2}+1\right)}
\end{equation}

\subsection{Calculations for next iterations \texorpdfstring{$(k > 1)$}{} in Grover's Search Algorithm with the second framework}\label{appendix: switched_kgt1}
After the first iteration, the switched depolarizing channel's output in block form,\\ \begin{equation}
    \rho_{\omega, 1}(1) = \mathcal{S}(\mathcal{D}_{\sqrt{t}}, \mathcal{D}_{\sqrt{t}})(\rho(1)\otimes\rho_c) = \begin{bmatrix}\theta \mathcal{F}(\rho(1))& \sqrt{\theta\overline{\theta}} \mathcal{R}(\rho(1))\\        \sqrt{\theta\overline{\theta}}\mathcal{R}(\rho(1)) & \overline{\theta} \mathcal{F} (\rho(1))\end{bmatrix}.
\end{equation}
Now, this is input for the next iteration, $k = 2$\\
\begin{eqnarray}
    \Bigl(\mathcal{G}\otimes\mathbb{I}_d\Bigl)\rho_{\omega, 1}(1)\Bigl(\mathcal{G}\otimes\mathbb{I}_d\Bigl)^{\dagger}=&\left(\mathcal{G}\otimes\begin{bmatrix}1 & 0\\0 & 1\end{bmatrix}\right)\rho_{\omega, 1}(1)\left(\mathcal{G}^{\dagger}\otimes\begin{bmatrix}1 & 0\\0 & 1\end{bmatrix}\right)\nonumber\\
    =&\begin{bmatrix}\mathcal{G} & 0\\0 & \mathcal{G}\end{bmatrix} \rho(1)  \begin{bmatrix}\mathcal{G}^{\dagger} &0\\0& \mathcal{G}^{\dagger}\end{bmatrix}\nonumber\\
    =&\begin{bmatrix}\mathcal{G} & 0 \\0 & \mathcal{G}\end{bmatrix} \begin{bmatrix}\theta \mathcal{F}(\rho(1))& \sqrt{\theta\overline{\theta}} \mathcal{R}(\rho(1))\\        \sqrt{\theta\overline{\theta}}\mathcal{R}(\rho(1)) & \overline{\theta} \mathcal{F} (\rho(1))\end{bmatrix}  \begin{bmatrix}\mathcal{G}^{\dagger} & 0\\0& \mathcal{G}^{\dagger}\end{bmatrix}\nonumber\\
    =&\begin{bmatrix}\theta \mathcal{G} \mathcal{F} (\rho(1))\mathcal{G}^{\dagger} & \sqrt{\theta\overline{\theta}} \mathcal{G} \mathcal{R} (\rho(1))\mathcal{G}^{\dagger}\\ \sqrt{\theta\overline{\theta}} \mathcal{G} \mathcal{R} (\rho(1))\mathcal{G}^{\dagger} & \overline{\theta}\mathcal{G} \mathcal{F} (\rho(1))\mathcal{G}^{\dagger}\end{bmatrix}\nonumber\\
    =& \begin{bmatrix}\theta \mathcal{F}(\rho(2))& \sqrt{\theta\overline{\theta}} \mathcal{R}(\rho(2))\\ \sqrt{\theta\overline{\theta}}\mathcal{R}(\rho(2)) & \overline{\theta} \mathcal{F} (\rho(2))\end{bmatrix} = \rho_{\omega, 1}(2)
\end{eqnarray}
\begin{eqnarray}\because\qquad, 
    \mathcal{G} \mathcal{F}(\rho(1))\mathcal{G}^{\dagger} 
    &&= f_{\rho} \mathcal{G}\rho(1)\mathcal{G}^{\dagger} + f_{\mathbb{I}} \mathcal{G}\mathrm{Tr}[\rho(1)]\dfrac{\mathbb{I}_d}{d}\mathcal{G}^{\dagger}
    = f_{\rho}\rho(2) + f_{\mathbb{I}}\dfrac{\mathbb{I}_d}{d} 
    = \underline{\mathcal{F}\bigl(\rho(2)\bigl)}\\
    \mathcal{G}\mathcal{R} (\rho(1))\mathcal{G}^{\dagger} 
    &&= r_{\rho}\mathcal{G}\rho(1)\mathcal{G}^{\dagger}+r_{\mathbb{I}}\mathcal{G}\mathrm{Tr}[\rho(1)]\dfrac{\mathbb{I}_d}{d}\mathcal{G}^{\dagger}
    = r_{\rho}\rho(2) + r_{\mathbb{I}}\dfrac{\mathbb{I}_d}{d} 
    = \underline{\mathcal{R}\bigl(\rho(2)\bigl)}
\end{eqnarray}

Now we apply the noise as $\bigl(D_{\sqrt{t}} \otimes \mathbb{I}_d\bigl)$ because the switch is noiseless. We can express the Kraus operators of the switch with two identical channels, $\bigl(D_{\sqrt{t}} \otimes \mathbb{I}_d\bigl)$, and the second control qubit $\rho_{c_2}$ as:\\
\begin{eqnarray}
    \mathcal{W}_{ij}^{(2)} &&= \Bigl(K_i \otimes \mathbb{I}_d\Bigl)\Bigl(K_j \otimes \mathbb{I}_d\Bigl) \otimes \ket{0}\bra{0} + \Bigl(K_i \otimes \mathbb{I}_d\Bigl)\Bigl(K_j \otimes \mathbb{I}_d\Bigl) \otimes \ket{1}\bra{1}\nonumber\\
    &&= \Bigl(K_iK_j\otimes\mathbb{I}_d\Bigl)\otimes\ket{0}\bra{0}+\Bigl(K_jK_i\otimes\mathbb{I}_d\Bigl)\otimes\ket{1}\bra{1}
\end{eqnarray}
\begin{equation}
    \rho_{\omega, 2}(2) = \mathcal{S}(D_{\sqrt{t}} , D_{\sqrt{t}})(\rho_{\omega, 1}(2) \otimes \rho_{c_2}) 
    = \mathlarger{\sum\limits}_{i, j =0}^{d^2}\mathcal{W}_{ij}^{(2)}\Bigl(\rho_{\omega, 1}(2) \otimes \rho_{c_2}\Bigl)\mathcal{W}_{ij}^{(2)\dagger}
\end{equation}
\begin{eqnarray*}
    =&& \mathlarger{\sum\limits}_{i, j = 0}^{d^2}\Bigl\{ \Bigl(K_iK_j \otimes \mathbb{I}_d\Bigl) \otimes \ket{0}\bra{0} + \Bigl(K_jK_i \otimes \mathbb{I}_d\Bigl) \otimes \ket{1}\bra{1} \Bigl\}\Bigl(\rho_{\omega, 1}(2) \otimes \rho_{c_2} \Bigl)\\
    &&\qquad \Bigl\{ \Bigl(K_iK_j \otimes \mathbb{I}_d\Bigl)^{\dagger} \otimes \ket{0}\bra{0} + \Bigl(K_jK_i \otimes \mathbb{I}_d\Bigl)^{\dagger} \otimes \ket{1}\bra{1} \Bigl\}\\
    =&& \mathlarger{\sum\limits}_{i,j = 0}^{d^2}\Bigl\{\Bigl(K_iK_j \otimes \mathbb{I}_d\Bigl)\rho_{\omega, 1}(2)\Bigl(K_iK_j \otimes \mathbb{I}_d\Bigl)^{\dagger} \otimes \ket{0}\bra{0}\rho_{c_2}\ket{0}\bra{0}\\
    &&\qquad +\Bigl(K_iK_j \otimes \mathbb{I}_d\Bigl)\rho_{\omega, 1}(2)\Bigl(K_jK_i \otimes \mathbb{I}_d\Bigl)^{\dagger} \otimes \ket{0}\bra{0}\rho_{c_2}\ket{1}\bra{1}\\
    &&\qquad +\Bigl(K_j K_i \otimes \mathbb{I}_d\Bigl)\rho_{\omega, 1}(2)\Bigl(K_i K_j \otimes \mathbb{I}_d\Bigl)^{\dagger} \otimes \ket{1}\bra{1}\rho_{c_2}\ket{0}\bra{0}\\
    &&\qquad +\Bigl(K_j K_i \otimes \mathbb{I}_d\Bigl)\rho_{\omega, 1}(2)\Bigl(K_j K_i \otimes \mathbb{I}_d\Bigl)^{\dagger} \otimes \ket{1}\bra{1}\rho_{c_2}\ket{1}\bra{1}\Bigl\}\\
\end{eqnarray*}
\begin{eqnarray}
    &&= \mathlarger{\mathlarger{\sum\limits}}_{i, j = 0}^{d^2} \Biggl\{ 
        \begin{bmatrix}K_iK_j & 0 \\ 0 & K_iK_j\end{bmatrix}
        \begin{bmatrix}\theta \mathcal{F}(\rho(2)) & \sqrt{\theta\overline{\theta}} \mathcal{F(\rho(2))} \\ \sqrt{\theta\overline{\theta}} \mathcal{R}(\rho(2)) & \overline{\theta} \mathcal{F}(\rho(2))\end{bmatrix}
        \begin{bmatrix}K_j^{\dagger}K_i^{\dagger} & 0 \\ 0 & K_j^{\dagger}K_i^{\dagger}\end{bmatrix}
    \otimes \theta\ket{0}\bra{0}\nonumber\\
    &&\qquad+ 
        \begin{bmatrix} K_iK_j & 0 \\ 0 & K_iK_j\end{bmatrix}
        \begin{bmatrix}\theta \mathcal{F}(\rho(2))                        & \sqrt{\theta\overline{\theta}} \mathbf{R}(\rho(2))\\\sqrt{\theta\overline{\theta}}\mathcal{R}(\rho(2)) & \overline{\theta}\mathcal{F} (\rho(2))\end{bmatrix}  
        \begin{bmatrix}K_i^{\dagger}K_j^{\dagger}&0\\0&K_i^{\dagger}K_j^{\dagger}\end{bmatrix} 
    \otimes \sqrt{\theta\overline{\theta}}\ket{0}\bra{1}\nonumber\\
    &&\qquad+ 
        \begin{bmatrix} K_jK_i & 0 \\ 0 & K_jK_i\end{bmatrix}
        \begin{bmatrix}\theta \mathcal{F}(\rho(2)) & \sqrt{\theta\overline{\theta}} \mathcal{F(\rho(2))} \\ \sqrt{\theta\overline{\theta}} \mathcal{R}(\rho(2)) & \overline{\theta} \mathcal{F}(\rho(2))\end{bmatrix}  
        \begin{bmatrix} K_j^{\dagger}K_i^{\dagger} & 0 \\ 0 & K_j^{\dagger}K_i^{\dagger}\end{bmatrix} 
    \otimes \sqrt{\theta\overline{\theta}}\ket{1}\bra{0}\nonumber\\
    &&\qquad +
        \begin{bmatrix}K_jK_i & 0 \\ 0 & K_jK_i\end{bmatrix}
        \begin{bmatrix}\theta \mathcal{F}(\rho(2)) & \sqrt{\theta\overline{\theta}} \mathcal{F(\rho(2))} \\ \sqrt{\theta\overline{\theta}} \mathcal{R}(\rho(2)) & \overline{\theta} \mathcal{F}(\rho(2))\end{bmatrix}
        \begin{bmatrix} K_i^{\dagger}K_j^{\dagger} & 0 \\ 0 & K_i^{\dagger}K_j^{\dagger}\end{bmatrix} 
    \otimes \overline{\theta}\ket{1}\bra{1}\Biggl\}\nonumber\\
    &&=\mathlarger{\mathlarger{\sum\limits}}_{i, j = 0}^{d^2} \Biggl\{\begin{bmatrix}\theta K_iK_j \mathcal{F}(\rho(2)) K_j^{\dagger}K_i^{\dagger} & \sqrt{\theta\overline{\theta}} K_iK_j \mathcal{R}(\rho(2)) K_j^{\dagger}K_i^{\dagger} \\ \sqrt{\theta\overline{\theta}}K_iK_j\mathcal{R}(\rho(2)) K_j^{\dagger}K_i^{\dagger} & \overline{\theta} K_iK_j\mathcal{F}(\rho(2))K_j^{\dagger}K_i^{\dagger}\end{bmatrix}  \otimes \theta\ket{0}\bra{0}\nonumber\\
    &&\qquad+ \begin{bmatrix}\theta K_iK_j \mathcal{F}(\rho_(2))K_i^{\dagger}K_j^{\dagger} & \sqrt{\theta\overline{\theta}} K_iK_j \mathcal{R}(\rho_(2))K_i^{\dagger}K_j^{\dagger} \\ \sqrt{\theta\overline{\theta}} K_iK_j \mathcal{R}(\rho_(2)) K_i^{\dagger}K_j^{\dagger} & \overline{\theta} K_iK_j \mathcal{F} (\rho_(2)) K_i^{\dagger}K_j^{\dagger}\end{bmatrix} 
    \otimes \sqrt{\theta\overline{\theta}}\ket{0}\bra{1}\nonumber\\
    &&\qquad+ \begin{bmatrix}\theta K_jK_i \mathcal{F}(\rho(2))K_j^{\dagger}K_i^{\dagger} & \sqrt{\theta\overline{\theta}} K_jK_i \mathcal{R}(\rho(2))K_j^{\dagger}K_i^{\dagger} \\ \sqrt{\theta\overline{\theta}}K_jK_i \mathcal{R}(\rho(2)) K_j^{\dagger}K_i^{\dagger} & \overline{\theta} K_jK_i \mathcal{F}(\rho(2))K_j^{\dagger}K_i^{\dagger}\end{bmatrix} \otimes \sqrt{\theta\overline{\theta}}\ket{1}\bra{0}\nonumber\\
    &&\qquad+\begin{bmatrix}\theta K_jK_i \mathcal{F}(\rho(2))K_i^{\dagger}K_j^{\dagger} & \sqrt{\theta\overline{\theta}} K_jK_i \mathcal{R}(\rho(2))K_i^{\dagger}K_j^{\dagger} \\ \sqrt{\theta\overline{\theta}}K_jK_i \mathcal{R}(\rho(2)) K_i^{\dagger}K_j^{\dagger} & \overline{\theta} K_jK_i \mathcal{F}(\rho(2))K_i^{\dagger}K_j^{\dagger}\end{bmatrix}
    \otimes\overline{\theta}\ket{1}\bra{1}\Biggl\} \label{eqn: kgt1_expression}
\end{eqnarray}
Eqn.\ref{eqn: kgt1_expression} can be written in block matrix form as:
\begin{equation}
    \rho_{\omega, 2}(2) = \begin{bmatrix} A_{00} & A_{01} \\ A_{10} & A_{11} \end{bmatrix}
\end{equation}
\noindent And for calculating each of the matrix elements in this block matrix, we can divide this summation into four parts as before: 
I. $j = 0, i = 0$,\quad II. $i\neq 0, j = 0$,\quad III. $i = 0, j \neq 0$ and \quad IV. $i\neq 0, j \neq 0$. 

\begin{eqnarray*}
A_{00} =&& \Biggl\{\begin{bmatrix}
    \theta\sum\limits_{i,j = 1}^{d^2} K_iK_j \mathcal{F}(\rho(2)) K_j^{\dagger}K_i^{\dagger} & \sqrt{\theta\overline{\theta}}\sum\limits_{i, j = 1}^{d^2} K_iK_j \mathcal{R}(\rho(2))K_j^{\dagger}K_i^{\dagger} \\ \sqrt{\theta\overline{\theta}}\sum\limits_{i, j = 1}^{d^2}K_iK_j \mathcal{R}(\rho(2)) K_j^{\dagger}K_i^{\dagger} & \overline{\theta}\sum\limits_{i, j = 1}^{d^2} K_iK_j \mathcal{F}(\rho(2))K_j^{\dagger}K_i^{\dagger}\end{bmatrix}\\
    &&\qquad+\begin{bmatrix}\theta \sum\limits_{j = 1}^{d^2}K_0K_j \mathcal{F}(\rho(2)) K_j^{\dagger}K_0^{\dagger} & \sqrt{\theta\overline{\theta}}\sum\limits_{j = 1}^{d^2} K_0 K_j \mathcal{R}(\rho(2))K_j^{\dagger}K_0^{\dagger} \\ \sqrt{\theta\overline{\theta}}\sum\limits_{j = 1}^{d^2}K_0 K_j \mathcal{R}(\rho(2)) K_j^{\dagger}K_0^{\dagger} & \overline{\theta}\sum\limits_{j = 1}^{d^2} K_0K_j \mathcal{F}(\rho(2))K_j^{\dagger}K_0^{\dagger}\end{bmatrix}\\
    &&\qquad+\begin{bmatrix}\theta \sum\limits_{i = 1}^{d^2}K_iK_0 \mathcal{F}(\rho(2))K_0^{\dagger}K_i^{\dagger} & \sqrt{\theta\overline{\theta}}\sum\limits_{i = 1}^{d^2} K_iK_0 \mathcal{R}(\rho(2))K_0^{\dagger}K_i^{\dagger}\\ \sqrt{\theta\overline{\theta}}\sum\limits_{i = 1}^{d^2}K_iK_0\mathcal{R}(\rho(2)) K_0^{\dagger}K_i^{\dagger} & \overline{\theta}\sum\limits_{i = 1}^{d^2} K_i K_0\mathcal{F}(\rho(2))K_0^{\dagger}K_i^{\dagger}\end{bmatrix}\\
    &&\qquad+\begin{bmatrix}\theta K_0K_0 \mathcal{F}(\rho(2))K_0^{\dagger}K_0^{\dagger} & \sqrt{\theta\overline{\theta}} K_0K_0 \mathcal{R}(\rho(2))K_0^{\dagger}K_0^{\dagger}\\ \sqrt{\theta\overline{\theta}}K_0K_0\mathcal{R}(\rho(2)) K_0^{\dagger}K_0^{\dagger} & \overline{\theta} K_0K_0\mathcal{F}(\rho(2))K_0^{\dagger}K_0^{\dagger}\end{bmatrix} \Biggl\}\;\otimes\;\theta\ket{0}\bra{0}\\
\end{eqnarray*}
\begin{eqnarray}
A_{00} =&& \Biggl\{(1-\sqrt{t}\Bigl)^2\begin{bmatrix}
    \theta\mathrm{Tr}[\mathcal{F}(\rho(2)]\dfrac{\mathbb{I}_d}{d} & \sqrt{\theta\overline{\theta}}\mathrm{Tr} [\mathcal{R}(\rho(2)]\dfrac{\mathbb{I}_d}{d} \\ \sqrt{\theta\overline{\theta}}\mathrm{Tr}[\mathcal{R}(\rho(2)]\dfrac{\mathbb{I}_d}{d} & \overline{\theta}\mathrm{Tr}[\mathcal{F}(\rho(2)]\dfrac{\mathbb{I}_d}{d}\end{bmatrix} 
    + t\begin{bmatrix}\theta \mathcal{F} (\rho(2))\dfrac{\mathbb{I}_d}{d} & \sqrt{\theta\overline{\theta}} \mathcal{R}\rho(2)) \dfrac{\mathbb{I}_d}{d} \\ \sqrt{\theta\overline{\theta}}\mathcal{R}(\rho(2)) \dfrac{\mathbb{I}_d}{d} & \overline{\theta} \mathcal{F}(\rho(2))\dfrac{\mathbb{I}_d}{d}\end{bmatrix}\nonumber\\
    &&\quad+ 2\sqrt{t}\Bigl(1-\sqrt{t}\Bigl)\begin{bmatrix}\theta \mathrm{Tr}[\mathcal{F}(\rho(2))]\dfrac{\mathbb{I}_d}{d} & \sqrt{\theta\overline{\theta}}\mathrm{Tr} [\mathcal{R}(\rho(2))]\dfrac{\mathbb{I}_d}{d} \\ \sqrt{\theta\overline{\theta}}\mathrm{Tr}[\mathcal{R}(\rho(2))]\dfrac{\mathbb{I}_d}{d} & \overline{\theta}\mathrm{Tr}[\mathcal{F}(\rho(2))]\dfrac{\mathbb{I}_d}{d}\end{bmatrix}\Biggl\}\; \otimes \; \theta\ket{0}\bra{0} \label{eqn: kgt1_A_00}
\end{eqnarray}
\begin{eqnarray*}
A_{01} =&& \Biggl\{\begin{bmatrix}
    \theta\sum\limits_{i, j = 1}^{d^2} K_iK_j \mathcal{F}(\rho(2)) K_i^{\dagger}K_j^{\dagger} & \sqrt{\theta\overline{\theta}}\sum\limits_{i, j = 1}^{d^2} K_iK_j \mathcal{R}(\rho(2))K_i^{\dagger}K_j^{\dagger} \\ \sqrt{\theta\overline{\theta}}\sum\limits_{i, j = 1}^{d^2}K_iK_j \mathcal{R}(\rho(2)) K_i^{\dagger}K_j^{\dagger} & \overline{\theta}\sum\limits_{i, j = 1}^{d^2} K_iK_j\mathcal{F}(\rho(2))K_i^{\dagger}K_j^{\dagger}\end{bmatrix}\\
    &&\qquad+ \begin{bmatrix}\theta \sum\limits_{j =1}^{d^2}K_0K_j\mathcal{F}(\rho(2)) K_0^{\dagger}K_j^{\dagger} & \sqrt{\theta\overline{\theta}}\sum\limits_{j = 1}^{d^2} K_0K_j \mathcal{R}(\rho(2))K_0^{\dagger}K_j^{\dagger} \\ \sqrt{\theta\overline{\theta}}\sum\limits_{j = 1}^{d^2}K_0K_jR(\rho(2)) K_0^{\dagger}K_j^{\dagger} & \overline{\theta}\sum\limits_{j = 1}^{d^2} K_0K_j\mathcal{F}(\rho(2))K_0^{\dagger}K_j^{\dagger}\end{bmatrix}\\
    &&\qquad+\begin{bmatrix}\theta \sum\limits_{i = 1}^{d^2}K_iK_0\mathcal{F}(\rho(2))K_i^{\dagger}K_0^{\dagger} & \sqrt{\theta\overline{\theta}}\sum\limits_{i = 1}^{d^2} K_iK_0 \mathcal{R}(\rho(2))K_i^{\dagger}K_0^{\dagger} \\ \sqrt{\theta\overline{\theta}}\sum\limits_{i = 1}^{d^2}K_iK_0\mathcal{R}(\rho(2))K_i^{\dagger}K_0^{\dagger} & \overline{\theta}\sum\limits_{i = 1}^{d^2} K_iK_0\mathcal{F}(\rho(2))K_i^{\dagger}K_0^{\dagger}\end{bmatrix}\\
    &&\qquad+\begin{bmatrix}\theta K_0K_0 \mathcal{F}(\rho(2))K_0^{\dagger}K_0^{\dagger} & \sqrt{\theta\overline{\theta}} K_0K_0 \mathcal{R}(\rho(2))K_0^{\dagger}K_0^{\dagger} \\ \sqrt{\theta\overline{\theta}}K_0K_0\mathcal{R}(\rho(2))K_0^{\dagger}K_0^{\dagger} & \overline{\theta} K_0K_0 \mathcal{F}(\rho(2))K_0^{\dagger}K_0^{\dagger}\end{bmatrix}\Biggl\}\;\otimes\;\sqrt{\theta\overline{\theta}}\ket{0}\bra{1}\}\\
\end{eqnarray*}
\begin{eqnarray}
A_{01} =&& \Biggl\{\Bigl(1-\sqrt{t}\Bigl)^2\begin{bmatrix}
    \theta\dfrac{\mathcal{F}(\rho(2))}{d^2} & \sqrt{\theta\overline{\theta}}\dfrac{\mathcal{R}(\rho(2))}{d^2} \\\sqrt{\theta\overline{\theta}}\dfrac{\mathcal{R}(\rho(2))}{d^2} \dfrac{\mathbb{I}_d}{d} & \overline{\theta}\dfrac{\mathcal{F}(\rho(2))}{d^2}\end{bmatrix}
   +t\begin{bmatrix}
    \theta \mathcal{F}(\rho(2)) & \sqrt{\theta\overline{\theta}}\mathcal{R}(\rho(2)) \\
    \sqrt{\theta\overline{\theta}}\mathcal{R}(\rho(2)) & \overline{\theta} \mathcal{F}(\rho(2))\end{bmatrix}\nonumber\\
    &&+ 2\sqrt{t}\Bigl(1-\sqrt{t}\Bigl)\begin{bmatrix}\theta \mathrm{Tr} [\mathcal{F}(\rho(2))]\dfrac{\mathbb{I}_d}{d} & \sqrt{\theta\overline{\theta}}\mathrm{Tr}[\mathcal{R}(\rho(2))]\dfrac{\mathbb{I}_d}{d} \\\sqrt{\theta\overline{\theta}}\mathrm{Tr}[\mathcal{R}(\rho(2))] \dfrac{\mathbb{I}_d}{d} & \overline{\theta}\mathrm{Tr}[\mathcal{F}(\rho(2))]\dfrac{\mathbb{I}_d}{d}\end{bmatrix}\Biggl\} \otimes \sqrt{\theta\overline{\theta}}\ket{0}\bra{1}\label{eqn: kgt1_A_01}
\end{eqnarray}
Other matrix elements $A_{10}$ and $A_{11}$ will also be similarly expanded as $A_{01}$ in \ref{eqn: kgt1_A_01} and $A_{00}$ in \ref{eqn: kgt1_A_00}, respectively, (eqn. except for the coefficients of the switch. Hence, \ref{eqn: kgt1_expression} can be expanded as:
\begin{eqnarray}
    &(1-\sqrt{t})^2&\Biggl\{\begin{bmatrix}\theta \mathrm{Tr}[\mathcal{F}(\rho(2))]\frac{\mathbb{I}_d}{d} & \sqrt{\theta\overline{\theta}} \mathrm{Tr}[\mathcal{R}(\rho(2))]\frac{\mathbb{I}_d}{d} \\ \sqrt{\theta\overline{\theta}} \mathrm{Tr}[\mathcal{R}(\rho(2))]\frac{\mathbb{I}_d}{d} & \overline{\theta} \mathrm{Tr}[\mathcal{F}(\rho(2))]\frac{\mathbb{I}_d}{d}\end{bmatrix} \otimes \theta\ket{0}\bra{0}\nonumber\\
    && \begin{bmatrix}\theta \frac{\mathcal{F}(\rho(2))}{d^2} & \sqrt{\theta\overline{\theta}} \frac{\mathcal{R}(\rho(2))}{d^2} \\ \sqrt{\theta\overline{\theta}} \frac{\mathcal{R}(\rho(2))}{d^2} & \overline{\theta} \frac{\mathcal{F}(\rho(2))}{d^2}\end{bmatrix} \otimes \sqrt{\theta\overline{\theta}}\ket{0}\bra{1}\nonumber\\
    && \begin{bmatrix}\theta \frac{\mathcal{F}(\rho(2))}{d^2} & \sqrt{\theta\overline{\theta}} \frac{\mathcal{R}(\rho(2))}{d^2} \\ \sqrt{\theta\overline{\theta}} \frac{\mathcal{R}(\rho(2))}{d^2} & \overline{\theta} \frac{\mathcal{F}(\rho(2))}{d^2} \end{bmatrix} \otimes \sqrt{\theta\overline{\theta}}\ket{1}\bra{0}\nonumber\\
    && \begin{bmatrix}\theta\mathrm{Tr}[\mathcal{F}(\rho(2))]\frac{\mathbb{I}_d}{d} & \sqrt{\theta\overline{\theta}}\mathrm{Tr}[\mathcal{R}(\rho(2))]\frac{\mathbb{I}_d}{d}\\ \sqrt{\theta\overline{\theta}} \mathrm{Tr}[\mathcal{R}(\rho(2))]\frac{\mathbb{I}_d}{d} & \overline{\theta} \mathrm{Tr}[\mathcal{F}(\rho(2))]\frac{\mathbb{I}_d}{d}\end{bmatrix} \otimes \overline{\theta}\ket{1}\bra{1}\Biggl\}\nonumber\\
    &+ 2\sqrt{t}(1-\sqrt{t})&\begin{bmatrix}\theta \mathrm{Tr}[\mathcal{F}(\rho(2)))]\frac{\mathbb{I}_d}{d} & \sqrt{\theta\overline{\theta}}\mathrm{Tr}[\mathcal{R}(\rho(2)))]\frac{\mathbb{I}_d}{d}\\ \sqrt{\theta\overline{\theta}} \mathrm{Tr}[\mathcal{R}(\rho(2)))]\frac{\mathbb{I}_d}{d} & \overline{\theta} \mathrm{Tr}[\mathcal{F}(\rho(2)))]\frac{\mathbb{I}_d}{d}\end{bmatrix} 
    \otimes \left(\theta\ket{0}\bra{0} + \sqrt{\theta\overline{\theta}}(\ket{0}\bra{1} + \ket{1}\bra{0}) + \overline{\theta}\ket{1}\bra{1}\right) \nonumber\\
    &+ t &\begin{bmatrix} \theta \mathcal{F}(\rho(2)) & \sqrt{\theta\overline{\theta}}\mathcal{R}(\rho(2)) \\ \sqrt{\theta\overline{\theta}}\mathcal{R}(\rho(2)) & \overline{\theta}\mathcal{F}(\rho(2))\end{bmatrix} \otimes \left(\theta\ket{0}\bra{0} + \sqrt{\theta\overline{\theta}}(\ket{0}\bra{1} + \ket{1}\bra{0}) + \overline{\theta}\ket{1}\bra{1}\right)
\end{eqnarray}
\begin{eqnarray*}
   =&&(1-\sqrt{t})^2 \Biggl\{ \begin{bmatrix} \theta \mathrm{Tr}[\mathcal{F}(\rho(2))]\frac{\mathbb{I}_d}{d} & \sqrt{\theta\overline{\theta}} \mathrm{Tr}[\mathcal{R}(\rho(2))]\frac{\mathbb{I}_d}{d} \\ \sqrt{\theta\overline{\theta}} \mathrm{Tr}[\mathcal{R}(\rho(2))]\frac{\mathbb{I}_d}{d} & \overline{\theta} \mathrm{Tr}[\mathcal{F}(\rho(2))]\frac{\mathbb{I}_d}{d}\end{bmatrix} \otimes \left(\theta\ket{0}\bra{0} + \overline{\theta}\ket{1}\bra{1}\right)\\
    &&\quad\quad\qquad +\begin{bmatrix}\theta \frac{\mathcal{F}(\rho(2))}{d^2} & \sqrt{\theta\overline{\theta}} \frac{\mathcal{R}(\rho(2))}{d^2}\\ \sqrt{\theta\overline{\theta}} \frac{\mathcal{R}(\rho(2))}{d^2} & \overline{\theta} \frac{\mathcal{F}(\rho(2))}{d^2} \end{bmatrix} \otimes \sqrt{\theta\overline{\theta}}\left(\ket{0}\bra{1} + \ket{1}\bra{0}\right)\Biggl\}\\
    &&\quad + \Biggl\{2\sqrt{t}(1-\sqrt{t}) \begin{bmatrix}\theta \mathrm{Tr}[\mathcal{F}(\rho(2))]\frac{\mathbb{I}_d}{d} & \sqrt{\theta\overline{\theta}} \mathrm{Tr}[\mathcal{R}(\rho(2))]\frac{\mathbb{I}_d}{d}\\ \sqrt{\theta\overline{\theta}} \mathrm{Tr}[\mathcal{R}(\rho(2))]\frac{\mathbb{I}_d}{d} & \overline{\theta} \mathrm{Tr}[\mathcal{F}(\rho(2))]\frac{\mathbb{I}_d}{d}\end{bmatrix} 
    + t \begin{bmatrix} \theta \mathcal{F}(\rho(2)) & \sqrt{\theta\overline{\theta}}\mathcal{R}(\rho(2))\\ \sqrt{\theta\overline{\theta}}\mathcal{R}(\rho(2)) & \overline{\theta}\mathcal{F}(\rho(2))\end{bmatrix}\Biggl\} 
    \otimes \rho_{c_2}
\end{eqnarray*}
Similar to $k = 1$, collecting terms in the above result and
following a similar substitution/notation eqn.(\ref{def: f(t)r(t)_notation}) as $k=1$,
\begin{eqnarray}
    =&&\left\{f_{\rho} \begin{bmatrix}\theta \mathcal{F}(\rho(2)) & \sqrt{\theta\overline{\theta}}\mathcal{R}(\rho(2))\\ \sqrt{\theta\overline{\theta}}\mathcal{R}(\rho(2)) & \overline{\theta}\mathcal{F}(\rho(2))\end{bmatrix} 
    + f_{\mathbb{I}} \begin{bmatrix}\theta \mathrm{Tr}[\mathcal{F}(\rho(2))]\frac{\mathbb{I}_d}{d} & \sqrt{\theta\overline{\theta}} \mathrm{Tr}[\mathcal{R}(\rho(2))]\frac{\mathbb{I}_d}{d}\\ \sqrt{\theta\overline{\theta}} \mathrm{Tr}[\mathcal{R}(\rho(2))]\frac{\mathbb{I}_d}{d} & \overline{\theta} \mathrm{Tr}[\mathcal{F}(\rho(2))]\frac{\mathbb{I}_d}{d} \end{bmatrix} \right\}\nonumber\\
    &&\qquad\otimes \left(\theta\ket{0}\bra{0} + \overline{\theta}\ket{1}\bra{1}\right)\nonumber\\
    &&\quad+ \left\{ r_{\rho} \begin{bmatrix}\theta \mathcal{F}(\rho(2)) & \sqrt{\theta\overline{\theta}}\mathcal{R}(\rho(2))\\ \sqrt{\theta\overline{\theta}}\mathcal{R}(\rho(2)) & \overline{\theta}\mathcal{F}(\rho(2)) \end{bmatrix} 
    + r_{\mathbb{I}} \begin{bmatrix} \theta \mathrm{Tr}[\mathcal{F}(\rho(2))]\frac{\mathbb{I}_d}{d} & \sqrt{\theta\overline{\theta}} \mathrm{Tr}[\mathcal{R}(\rho(2))]\frac{\mathbb{I}_d}{d}\\ \sqrt{\theta\overline{\theta}} \mathrm{Tr}[\mathcal{R}(\rho(2))]\frac{\mathbb{I}_d}{d} & \overline{\theta} \mathrm{Tr}[\mathcal{F}(\rho(2))]\frac{\mathbb{I}_d}{d}\end{bmatrix} \right\}\nonumber\\ 
    &&\qquad\otimes \sqrt{\theta\overline{\theta}}\left(\ket{0}\bra{1} + \ket{1}\bra{0}\right)    
    \end{eqnarray}
\noindent Comparing this equation with the notations in eqn.(\ref{eqn: k=1_result}) and eqn.(\ref{def: F(rho)R(rho)_notation})
\begin{eqnarray*}
    =&&\Biggl\{ \underbrace{ 
    f_{\rho} \begin{bmatrix} \theta \mathcal{F}(\rho(2)) & \sqrt{\theta\overline{\theta}}\mathcal{R}(\rho(2))\\ \sqrt{\theta\overline{\theta}}\mathcal{R}(\rho(2)) & \overline{\theta}\mathcal{F}(\rho(2)) \end{bmatrix}}
    _{f_{\rho}(.) + } + 
    \underbrace{ f_{\mathbb{I}} \begin{bmatrix} \theta \mathrm{Tr}[\mathcal{F}(\rho(2))] & \sqrt{\theta\overline{\theta}} \mathrm{Tr}[\mathcal{R}(\rho(2))]\\ \sqrt{\theta\overline{\theta}} \mathrm{Tr}[\mathcal{R}(\rho(2))] & \overline{\theta} \mathrm{Tr}[\mathcal{F}(\rho(2))]\end{bmatrix} \frac{\mathbb{I}_d}{d}}
    _{f_{\mathbb{I}} \mathrm{Tr}_{d \times d}(.) \frac{\mathbb{I}_d}{d}}\Biggl\}\\
    &&\qquad\otimes \left(\theta\ket{0}\bra{0} + \overline{\theta}\ket{1}\bra{1}\right)\\
    &&\quad+ \Biggl\{\underbrace{r_{\rho} \begin{bmatrix}\theta \mathcal{F}(\rho(2)) & \sqrt{\theta\overline{\theta}}\mathcal{R}(\rho(2))\\ \sqrt{\theta\overline{\theta}}\mathcal{R}(\rho(2)) & \overline{\theta}\mathcal{F}(\rho(2))\end{bmatrix}}
    _{r_{\rho}(.) + } + 
    \underbrace{r_{\mathbb{I}} \frac{\mathbb{I}_d}{d}\otimes\begin{bmatrix} \theta \mathrm{Tr}[\mathcal{F}(\rho(2))] & \sqrt{\theta\overline{\theta}} \mathrm{Tr}[\mathcal{R}(\rho(2))]\\ \sqrt{\theta\overline{\theta}} \mathrm{Tr}[\mathcal{R}(\rho(2))] & \overline{\theta} \mathrm{Tr}[\mathcal{F}(\rho(2))] \end{bmatrix}}
    _{r_{\mathbb{I}} \mathrm{Tr}_{d\times d}(.) \frac{\mathbb{I}_d}{d}} \Biggl\}\\ 
     &&\qquad\otimes \sqrt{\theta\overline{\theta}}\left(\ket{0}\bra{1} + \ket{1}\bra{0}\right)
\end{eqnarray*}
     \noindent Thus, eqn.(\ref{eqn: k=2_result}) is a $4d$ dimensional block matrix which is again structurally similar to eqn.(\ref{eqn: k=1_result}) and we can expand the previous notation of $\mathcal{F}$ and $\mathcal{R}$ to write the above result as:
\begin{eqnarray*}
    =&&\{f_{\rho}(\rho_{\omega, 1}(2)) + f_{\mathbb{I}} \mathrm{Tr}_{d\times d}[\rho_{\omega, 1}(2)] \frac{\mathbb{I}_d}{d}\}\otimes \left(\theta\ket{0}\bra{0} + \overline{\theta}\ket{1}\bra{1}\right)\\ 
    &&\quad+ \{r_{\rho}(\rho_{\omega, 1}(2)) + r_{\mathbb{I}} \mathrm{Tr}_{d\times d}[\rho_{\omega, 1}(2)] \frac{\mathbb{I}_d}{d}\}\otimes\sqrt{\theta\overline{\theta}}\left(\ket{0}\bra{1} + \ket{1}\bra{0}\right)\\
    =&&\mathcal{F}(\rho_{\omega, 1}(2))\otimes \left(\theta\ket{0}\bra{0} + \overline{\theta}\ket{1}\bra{1}\right) + \mathcal{R}(\rho_{\omega, 1}(2))\otimes\sqrt{\theta\overline{\theta}}\left(\ket{0}\bra{1} + \ket{1}\bra{0}\right)\\
\rho_{\omega, 2}(2)=&&\begin{bmatrix}
        \theta \mathcal{F}(\rho_{\omega, 1}(2)) & \sqrt{\theta\overline{\theta}} \mathcal{R}(\rho_{\omega, 1}(2)) \\ \sqrt{\theta\overline{\theta}} \mathcal{R}(\rho_{\omega, 1}(2)) & \overline{\theta} \mathcal{F}(\rho_{\omega, 1}(2))
    \end{bmatrix}
\end{eqnarray*}
\noindent Here, we use the shorthand notation introduced in eqn.(\ref{def: rho_omega1(1)}). The superscript denotes the number of switches correlated to the input state, and $\omega$ in the subscript signifies the second framework. The subscript also contains the number of iterations of Grover's Algorithm the input state has traversed through, 
\subsection{Find the state after \texorpdfstring{$k$}{} Grover iterations where the system is correlated with \texorpdfstring{$k$}{} switches
\texorpdfstring{$= \rho_{\omega, k}(k)$}{}}
\noindent We denote the state of the system after measurement after k iterations as $M_k$:
\begin{equation}\label{eqn: M_k-sub}
   M_k[\rho_{\omega, k}(k)] = (\mathbb{I}_d \otimes \bra{+}^{\otimes k}) \rho_{\omega, k}(k) (\mathbb{I}_d \otimes \ket{+}^{\otimes k}) 
\end{equation}
\noindent We have $\mathbb{I}_d$ as part of the measurement operation because we want to keep the input state intact and only trace out the quantum switches correlated with the input state. 
\begin{eqnarray*}
&&= \left((\mathbb{I}_d \otimes \bra{+}^{\otimes k-1}) \otimes \bra{+}\right) \rho_{\omega, k}(k) \left((\mathbb{I}_d \otimes \ket{+}^{\otimes k-1}) \otimes \ket{+}\right)\\ 
&&= \left((\mathbb{I}_d \otimes \bra{+}^{\otimes k-1}) \otimes \begin{bmatrix}
    \frac{1}{\sqrt{2}} & \frac{1}{\sqrt{2}}
\end{bmatrix}\right) \rho_{\omega, k}(k) \left((\mathbb{I}_d \otimes \ket{+}^{\otimes k-1}) \otimes 
\begin{bmatrix}
    \frac{1}{\sqrt{2}} \\ \frac{1}{\sqrt{2}}
\end{bmatrix}\right)\\
&&= \begin{bmatrix}
    \frac{\mathbb{I}_d \otimes \bra{+}^{\otimes k-1}}{\sqrt{2}} & \frac{\mathbb{I}_d \otimes \bra{+}^{\otimes k-1}}{\sqrt{2}}
\end{bmatrix} \begin{bmatrix}
    \theta \mathcal{F}(\rho_{\omega, k-1}(k)) & \sqrt{\theta\overline{\theta}} \mathcal{R}(\rho_{\omega, k-1}(k)) \\ \sqrt{\theta\overline{\theta}} \mathcal{R}(\rho_{\omega, k-1}(k)) & \overline{\theta} \mathcal{F}(\rho_{\omega, k-1}(k))
 \end{bmatrix} \begin{bmatrix}
    \frac{\mathbb{I}_d \otimes \bra{+}^{\otimes k-1}}{\sqrt{2}} \\ \frac{\mathbb{I}_d \otimes \bra{+}^{\otimes k-1}}{\sqrt{2}}
 \end{bmatrix}\\
 &&= \begin{bmatrix}
    (\mathbb{I}_d \otimes \bra{+}^{\otimes k-1})(\theta \mathcal{F}(\rho_{\omega, k-1}(k))) + (\mathbb{I}_d \otimes \bra{+}^{\otimes k-1}) (\sqrt{\theta\overline{\theta}} \mathcal{R}(\rho_{\omega, k-1}(k))) \\  (\mathbb{I}_d \otimes \bra{+}^{\otimes k-1}) (\sqrt{\theta\overline{\theta}} \mathcal{R}(\rho_{\omega, k-1}(k))) +  (\mathbb{I}_d \otimes \bra{+}^{\otimes k-1})(\overline{\theta} \mathcal{F}(\rho_{\omega, k-1}(k)))
 \end{bmatrix}^{\mathcal{T}}\begin{bmatrix}
    \mathbb{I}_d \otimes \ket{+}^{\otimes k-1} \\ \mathbb{I}_d \otimes \ket{+}^{\otimes k-1}
 \end{bmatrix}\\
&&= 
   (\mathbb{I}_d \otimes \bra{+}^{\otimes k-1})(\mathcal{F}(\rho_{\omega, k-1}(k)))(\mathbb{I}_d \otimes \ket{+}^{\otimes k-1}) 
   + 2\sqrt{\theta\overline{\theta}} (\mathbb{I}_d \otimes \bra{+}^{\otimes k-1}) (\mathcal{R}(\rho_{\omega, k-1}(k))) (\mathbb{I}_d \otimes \ket{+}^{\otimes k-1})
\end{eqnarray*}
Using eqn.(\ref{eqn: F_k-sub}) and eqn.(\ref{eqn: R_k-sub}) in the result above:
\begin{eqnarray}    
   \frac{1}{2}
   (\mathbb{I}_d \otimes \bra{+}^{\otimes k-1})\left(f_{\rho}(\rho_{\omega, k-1}(k)) + f_{\mathbb{I}}\frac{\mathbb{I}_d}{d}\otimes\mathrm{Tr}_{d \times d}[\rho_{\omega, k-1}(k)]\right)(\mathbb{I}_d \otimes \ket{+}^{\otimes k-1})\nonumber\\
   + 2\sqrt{\theta\overline{\theta}} (\mathbb{I}_d \otimes \bra{+}^{\otimes k-1})\left(r_{\rho}(\rho_{\omega, k-1}(k)) + r_{\mathbb{I}}\frac{\mathbb{I}_d}{d}\otimes\mathrm{Tr}_{d \times d}[\rho_{\omega, k-1}(k)]\right)(\mathbb{I}_d \otimes \ket{+}^{\otimes k-1}).
\end{eqnarray}
Rearranging the terms, we get:
\begin{eqnarray}
&\frac{1}{2}\left\{
   (f_{\rho} + 2\sqrt{\theta\overline{\theta}}r_{\rho})(\mathbb{I}_d \otimes \bra{+}^{\otimes k-1})(\rho_{\omega, k-1}(k))(\mathbb{I}_d \otimes \ket{+}^{\otimes k-1})\right.\nonumber\\
   &\left. +  (f_{\mathbb{I}}+ 2\sqrt{\theta\overline{\theta}}r_{\mathbb{I}})(\mathbb{I}_d \otimes \bra{+}^{\otimes k-1})(\frac{\mathbb{I}_d}{d}\otimes\mathrm{Tr}_{d \times d}[\rho_{\omega, k-1}(k)](\mathbb{I}_d \otimes \ket{+}^{\otimes k-1})
\right\}
\end{eqnarray}
Thus,
eqn. \ref{eqn: M_k-sub}
gives us
a recursive recipe to get the density operator after k iterations:
\begin{equation}
   M_{k}(\rho_{\omega, k}(k)) = \frac{1}{2}\left\{
   (f_{\rho} + 2\sqrt{\theta\overline{\theta}}r_{\rho})M_{k-1}(\rho_{\omega, k-1}(k)) 
   +  (f_{\mathbb{I}}+ 2\sqrt{\theta\overline{\theta}}r_{\mathbb{I}})(\mathbb{I}_d \otimes \bra{+}^{\otimes k-1}\mathrm{Tr}_{d \times d}[\rho_{\omega, k-1}(k)]\ket{+}^{\otimes k-1}\right\}
\end{equation}
After that, we can analyze the density operator obtained for success probability after k Grover iterations.
\end{document}